\documentclass[12pt]{iopart}
\usepackage{iopams,epsfig}

\newcommand{\er}{{\mathbb R}}
\newcommand{\en}{{\mathbb N}}

\newcommand{\boldx}{{\mathbf x}}
\newcommand{\dsigma}{\mbox{d}\sigma}

\newcommand{\dx}{\mbox{d}x}
\newcommand{\dy}{\mbox{d}y}
\newcommand{\dv}{\mbox{d}v}
\newcommand{\dxi}{\mbox{d}\xi}
\newcommand{\deta}{\mbox{d}\eta}
\newcommand{\dk}{\mbox{d}k}
\newcommand{\erf}{\mathop{\mbox{erf}}}

\begin{document}

\title[Boundary conditions for stochastic simulations]{Realistic
boundary conditions for stochastic simulations of reaction-diffusion processes}

\author{
Radek Erban and S. Jonathan Chapman
}

\address{University of Oxford, Mathematical Institute,
24-29 St. Giles', Oxford, OX1 3LB, United Kingdom}

\ead{erban@maths.ox.ac.uk; chapman@maths.ox.ac.uk}

\begin{abstract}
Many cellular and subcellular biological processes can be 
described in terms of diffusing and chemically reacting 
species (e.g. enzymes). Such reaction-diffusion processes 
can be mathematically modelled using either deterministic 
partial-differential equations or stochastic simulation 
algorithms. The latter provide a more detailed and precise 
picture, and several stochastic simulation algorithms have 
been proposed in recent years. Such models typically
give the same description of the reaction-diffusion 
processes far from the boundary of the simulated domain, 
but the behaviour close to a reactive boundary (e.g. 
a membrane with receptors) is unfortunately model-dependent.
In this paper, we study four different approaches to stochastic 
modelling of reaction-diffusion problems and show the 
correct choice of the boundary condition for each model.
The reactive boundary is treated as partially reflective,
which means that some molecules hitting the boundary
are adsorbed (e.g. bound to the receptor) and some
molecules are reflected. The probability that the molecule 
is adsorbed rather than reflected depends on the reactivity 
of the boundary (e.g. on the rate constant of the adsorbing 
chemical reaction and on the number of available receptors),
and on the stochastic model used. This dependence is derived
for each model. 
\end{abstract}




\section{Introduction}

Let us consider a system of $k$ chemicals diffusing and reacting
in a domain $\Omega \subset \er^3$. Let $n_j(\boldx,t)$, $j=1, \dots, k$, 
be the density of molecules of the $j$-th chemical species
at the point $\boldx \in \Omega$. Assuming that there are a lot of 
molecules present in the system, the time evolution of density 
$n_j(\boldx,t)$ can be computed by solving the system of reaction-diffusion 
partial-differential equations
\begin{equation}
\frac{\partial n_j}{\partial t}
=
D_j
\nabla^2 n_j
+
R_j(n_1,n_2,\dots,n_k),
\qquad
j=1, \dots, k,
\label{reactiondiffusionPDE}
\end{equation}
where $D_j$ is the diffusion constant of the $j$-th chemical species,
$\nabla^2$ is the Laplace operator and reaction term 
$R_j(n_1,n_2,\dots,n_k)$ takes into account the chemical reactions 
which modify the concentration of the $j$-th chemical species.
To describe uniquely the time evolution of the system, we have
to introduce suitable boundary conditions for
the system of equations (\ref{reactiondiffusionPDE}).
The simplest boundary conditions can be formulated
in terms of vanishing density $n_j(\boldx,t)$ on
the boundary of $\Omega$ or vanishing flux through the
boundary of $\Omega$. Coupling system of equations 
(\ref{reactiondiffusionPDE}) with such a boundary
condition, we can compute densities $n_j(\boldx,t)$ at
any time $t$ from the initial densities $n_j(\boldx,0),$
$j=1,\dots,k$.

Reaction-diffusion processes in biology often involve
low molecular abundancies of some chemical species.
In such a case, the continuum deterministic
description (\ref{reactiondiffusionPDE}) is no longer
valid and suitable stochastic models
must be used instead. Various stochastic simulation 
algorithms have been proposed in the literature 
\cite{Andrews:2004:SSC,Hattne:2005:SRD,Isaacson:2006:IDC,Stundzia:1996:SSC}.
In general, the stochastic treatment of diffusion and
first-order reactions (such as degradation or conversion)
is well understood. There is less understanding (and 
stochastic models differ) when second-order
chemical reactions are taken into account, e.g. reactions
in which two molecules collide for the reaction to take place. 
Another important
problem is the implementation of the correct boundary conditions
for the stochastic simulation algorithms. On the one hand, 
the simple boundary conditions mentioned above are easy to reformulate
in the stochastic case -- the vanishing density on the boundary
of $\Omega$ simply means that a diffusing molecule is
removed from the system whenever it hits the boundary;
and the vanishing flux through the boundary means that a diffusing 
molecule is reflected whenever it hits the boundary. On the
other hand, more realistic boundary conditions have to be
handled with care. They can be formulated in terms of the 
{\it partially adsorbing boundary}, which means that some molecules 
hitting the boundary are reflected and some are adsorbed. 
The partially adsorbing boundary corresponds to the so-called 
Robin boundary condition of the macroscopic 
partial-differential equation (\ref{reactiondiffusionPDE}). However,
this correspondence is model-dependent.

We will see later, in Section \ref{secrd}, that the derivation 
of the correct boundary condition depends on the stochastic model 
of the diffusion but not on the stochastic model of the chemical 
reactions in the solution. Consequently, we start this paper by 
studying stochastic models of diffusion only. 
In Section \ref{sec4models}, we introduce four different stochastic 
approaches to model molecular diffusion and we state
the appropriate boundary conditions. In Section \ref{illcomp},
we present illustrative simulations of all four models,
validating the boundary conditions presented. Moreover, we 
also clearly illustrate that the 
boundary conditions are indeed model-dependent. In Section
\ref{secmath}, we present the mathematical
derivation of the boundary conditions for each model,
i.e. we provide a theoretical justification of results from
Section \ref{sec4models}. Moreover,
we show that all four models are suitable for modelling
diffusion far from the reactive boundary. Section
\ref{secmath} is intended for a more theoretical
audience and can be skipped by a reader who is not interested
in the mathematical justification of the boundary conditions
and stochastic models. In Section \ref{secrd}, we show that 
reaction-diffusion models can be treated using the same boundary
conditions which were previously derived for the corresponding models 
of the diffusion only. We conclude with summary and outlook 
in Section \ref{summaryoutlook}.

\section{Boundary conditions for stochastic models of diffusion}

\label{sec4models}

The boundary condition of any stochastic simulation algorithm 
can be formulated as follows:  {\it whenever a molecule hits 
the boundary, it is adsorbed with some probability, and reflected 
otherwise}. The special cases of this boundary condition
are: (a) the molecule is always reflected (such a boundary is
called the reflecting boundary in what follows); and
(b) the molecule is always adsorbed (in this case the boundary
is called fully adsorbing). The reflecting boundary condition
is often used when no adsorption of the diffusing molecules 
on the boundary takes place. On the other hand, if the molecule 
can chemically or physically attach to the boundary, then one 
has to assume that  the boundary is (at least) partially adsorbing.

The important question is, what is the probability
that the particle is adsorbed rather than reflected, and how
does this relate to the reactive properties of the boundary
for a given stochastic model? To answer this question, let us 
follow the $x$-coordinate of the diffusing molecule (the other 
coordinates can be treated similarly), so that we 
study the diffusion of molecules in the one-dimensional 
interval $[0,L]$ where $L$ is the length of the computational
domain. Assuming that we have a lot of molecules in the
system, we can describe the system in terms of density $n(x,t)$ 
of molecules at point $x \in [0,L]$ and time $t$, so 
that $n(x,t) \, \delta x$ gives the number of molecules in
the small interval $[x,x+\delta x]$ at time $t$. The evolution
of $n(x,t)$ is governed by the diffusion equation 
\begin{equation}
\frac{\partial n}{\partial t}
=
D \frac{\partial^2 n}{\partial x^2},
\qquad
\mbox{for} \; x \in [0,L], \; t \ge 0,
\label{diffusionequation1D}
\end{equation}
where $D$ is the diffusion constant. The general first-order
reactive boundary condition
at $x=0$ is the so-called Robin boundary condition
\begin{equation}
D \, \frac{\partial n}{\partial x} (0,t)
=
K \, n(0,t)
\label{RobinBD}
\end{equation}
where the constant $K$ describes the reactivity of the boundary 
(see Appendix for the relation between $K$ and the
chemical properties of the boundary) and may in general 
depend on time.
The boundary condition at right boundary $x = L$ can be treated
similarly. 

In the following four subsections we introduce four stochastic
models of diffusion. 
The first model, introduced in Section \ref{secmod1des}, is a position 
jump process on a lattice. This model is discrete in both time and space, 
and is used in a stochastic simulation algorithm 
which is based on the reaction-diffusion master 
equation \cite{Hattne:2005:SRD, Isaacson:2006:IDC}.
The second model, introduced in Section
\ref{secmod2des}, is again discrete in time and discontinuous
in space but the positions of molecules are not confined to
a regular lattice. It is essentially 
the Euler scheme for the Smoluchowski stochastic 
differential equation, which is the core of the stochastic 
approach of Andrews and Bray \cite{Andrews:2004:SSC}.
The third scheme, introduced in Section \ref{secmod3des}, 
is a discrete velocity jump process which
is a discrete in time, continuous in space random walk 
with discretized velocities, where the velocities evolve 
on a finite lattice. The last stochastic model of diffusion,
introduced in Section \ref{secmod4des},
is the Euler scheme for the solution of the stochastic 
Langevin equation. 
It is a velocity jump process (i.e. a random walk discrete in time, 
continuous in space and discontinuous in velocities)
where the Brownian particle can move with any real value of
the velocity. In all four cases, we study the connections between 
the boundary conditions of the stochastic simulation and Robin
boundary condition (\ref{RobinBD}) of the macroscopic diffusion 
equation (\ref{diffusionequation1D}). We provide the relation 
between $K$ in (\ref{RobinBD}) and the parameters of each model.
The mathematical derivation of these relations is included
later, in Section \ref{secmath}.

\subsection{Position Jump Process I}

\label{secmod1des}

Let us discretize the domain of interest $[0,L]$ into $M$ lattice points 
a distance $h = L/M$ apart, namely we consider the lattice
\begin{equation}
\left\{
\frac{h}{2},
\frac{3h}{2},
\frac{5h}{2},
\frac{7h}{2},
\frac{9h}{2}.
\dots
\frac{(2M-1)h}{2}
\right\}.
\label{lattice}
\end{equation}
Let us choose time step $\Delta t$ such that $2 D \Delta t \ll h^2.$
We simulate a system of $N$ molecules whose positions
are assumed to be at one of the discrete positions (\ref{lattice}).
Let $x_i(t)$ be the position of the $i$-th molecule,
$i = 1, 2, \dots, N$, at time $t$. The position $x_i(t+\Delta t)$
is computed from the position $x_i(t)$ as follows:
\begin{equation}
x_i(t+\Delta t)
=
\cases{x_i(t)& with probability $1 - 2 D \Delta t/h^2$,\\
x_i(t) - h& with probability $D \Delta t/h^2$, \\
x_i(t) + h& with probability $D \Delta t/h^2$.\\}
\label{mod1step}
\end{equation}
At $x=0$, we implement the following boundary condition:
{\it whenever a molecule hits the boundary, it is 
adsorbed with probability $P_1 h$, and reflected
otherwise}. Here, $P_1$ is a given nonnegative constant. 
The implementation of this boundary condition at $x=0$ is performed
as follows. If the $i$-th molecule is at position 
$x_i(t)=h/2$ and attempts to jump to the left, then
\begin{equation}
x_i(t+\Delta t)
=
h/2
\qquad \mbox{with probability} \; 1 - P_1 h.
\label{partadsdiscrete}
\end{equation}
Otherwise, we remove the molecule from the system.  We show in 
Section \ref{secmod1math} that the random walk (\ref{mod1step})
with boundary condition (\ref{partadsdiscrete})
leads to the diffusion equation (\ref{diffusionequation1D}) 
with Robin boundary condition (\ref{RobinBD}), where 
\begin{equation}
K
=
P_1 D,
\quad
\mbox{which is equivalent to}
\quad
P_1
=
\frac{K}{D}.
\label{Kmod1}
\end{equation}

\subsection{Position Jump Process II}

\label{secmod2des}
 
Let us choose a time step $\Delta t$. Let $x_i(t)$, $i=1,2, \dots, N$, be 
the position of the $i$-th molecule at time $t$. 
The position $x_i(t+\Delta t)$ is computed from the position 
$x_i(t)$ as follows:
\begin{equation}
x_i(t + \Delta t) = x_i(t) + \sqrt{2 D \, \Delta t} \; \zeta_i,
\qquad i = 1, \dots, N,
\label{discretestochevol}
\end{equation}
where $\zeta_i$ is normally distributed random variable with zero
mean and unit variance. This random walk is essentially 
the Euler scheme for the Smoluchowski stochastic 
differential equation (\ref{SDEmod2}) as discussed later,
in Section \ref{secmod2math}.
We implement the following partially adsorbing boundary condition 
at $x=0$: {\it whenever a molecule hits the boundary, it is 
adsorbed with probability $P_2 \sqrt{\Delta t}$, and reflected
otherwise}. Obviously, if $x_i(t + \Delta t)$ computed by 
(\ref{discretestochevol}) is negative, a molecule has hit the boundary. 
However, Andrews and Bray \cite{Andrews:2004:SSC} 
argue that there is a chance that a molecule hit the boundary 
during the finite time step even if $x_i(t + \Delta t)$ 
computed by (\ref{discretestochevol}) is positive that is,
during the time interval $[t,t+\Delta t]$ the molecule might
have crossed to $x_i$ negative and then crossed back to 
$x_i$ positive again. They found
that the probability that the molecule hit the boundary
$x=0$ at least once during the time step  $\Delta t$
is  $\exp[- x_i(t) x_i(t + \Delta t)/(D \Delta t)]$
for $x_i(t) \ge 0$, $x_i(t + \Delta t) \ge 0$. Consequently, the partially
reflective boundary condition is implemented as follows:

\medskip
\leftskip 1cm
 
\noindent 
(a) If $x_i(t+\Delta t)$ computed by (\ref{discretestochevol}) 
is negative then 
$x_i(t+\Delta t) = - x_i(t) - \sqrt{2 D \, \Delta t} \; \zeta_i$
with probability $1 - P_2 \sqrt{\Delta t}$, otherwise
we remove the molecule from the system.

\smallskip

\noindent
(b) If $x_i(t+\Delta t)$ computed by  (\ref{discretestochevol}) is 
positive then we remove the molecule from the system
with probability $\exp[- x_i(t) x_i(t + \Delta t)/(D \Delta t)]
P_2 \sqrt{\Delta t}$.
 
\medskip
\leftskip 0cm

\noindent
The partially adsorbing boundary condition (a) - (b) leads to
the Robin boundary condition (\ref{RobinBD}) with 
\begin{equation}
K
=
\frac{2 P_2 \sqrt{D}}{\sqrt{\pi}},
\quad
\mbox{which is equivalent to}
\quad
P_2
=
\frac{K \sqrt{\pi}}{2 \sqrt{D}}.
\label{Kmod2}
\end{equation}
Let us note that some authors use the case (a) only as the implementation
of the partially reflective boundary condition \cite{Singer:2006:PRD},
i.e. they do not take the Andrews and Bray correction (b) into account.
Considering the random walk (\ref{discretestochevol}) with the 
boundary condition (a) only, the parameter $K$ of Robin boundary 
condition (\ref{RobinBD}) can be computed as
\begin{equation}
K
=
\frac{P_2 \sqrt{D}}{\sqrt{\pi}},
\quad
\mbox{which is equivalent to}
\quad
P_2
=
\frac{K \sqrt{\pi}}{\sqrt{D}}.
\label{Kmod2alt}
\end{equation}
Comparing (\ref{Kmod2}) and (\ref{Kmod2alt}), we see
that we lose a factor of 2 if we do not consider the Andrews 
and Bray correction (b). The mathematical
justification of formulas (\ref{Kmod2}) and (\ref{Kmod2alt})
is presented in Section \ref{secmod2math}.

\subsection{Velocity Jump Process I}

\label{secmod3des}

We consider that each molecule moves along the $x$-axis at a constant (large)
speed $s$, but that at random instants of time it reverses its direction 
according to a Poisson process with the turning frequency 
\begin{equation}
\lambda = \frac{s^2}{2D}\,.
\label{lambdasD}
\end{equation}
Therefore, the $i$-th molecule is described by two variables: its 
position $x_i(t)$ and its velocity $v_i(t) =  \pm s$. We use a small time
step $\Delta t$ such that $\lambda \Delta t \ll 1$.
During each time step a molecule moves with speed $s$ in the chosen
direction.  At the end of each time step, uniformly distributed
random number $r_i \in [0,1]$ is generated. If $r_i < \lambda \delta t$,
then the $i$-th molecule changes its direction, so that
it will move during the next time step in the opposite 
direction. 

We implement the following partially adsorbing boundary 
condition at $x=0$: {\it whenever a molecule hits the boundary, 
it is adsorbed with probability $P_3/s$, and reflected otherwise}.
Here, $P_3$ is a given nonnegative number. The partially adsorbing 
boundary condition at $x=0$ is implemented as follows. If
the position of the $i$-th molecule satisfies 
$x_i(t+\Delta t) < 0$ at the end of the time step, then 
\begin{equation}
\left.
\begin{array}{ccc}
x_i(t + \Delta t) & = & - x_i(t) - v_i(t) \Delta t  \\
v_i(t + \Delta t) & = & - v_i(t) 
\end{array}
\; \right\}
\quad \mbox{with probability} \; 1 - \frac{P_3}{s}\,,
\label{partadsveljump}
\end{equation}
otherwise, the $i$-th molecule is removed from the system.
It can be shown that this velocity jump process leads to diffusion 
equation (\ref{diffusionequation1D}) provided that $s$ is 
sufficiently large (see Section \ref{secmod3math} for details). 
Boundary condition (\ref{partadsveljump}) can be related with 
Robin boundary condition (\ref{RobinBD}), with  
\begin{equation}
K
=
\frac{P_3}{2},
\quad
\mbox{which is equivalent to}
\quad
P_3
=
2 K.
\label{Kmod3}
\end{equation}

\subsection{Velocity Jump Process II}

\label{secmod4des}

Let us choose time step $\Delta t$. The $i$-th molecule is described 
by two variables: its position $x_i(t)$ and its velocity $v_i(t)$.
We compute position $x_i(t+\Delta t)$ and velocity $v_i(t+\Delta t)$ 
from position $x_i(t)$ and velocity $v_i(t)$ by formulas
\begin{eqnarray}
x_i(t+\Delta t) & = & x_i(t) + v_i(t) \Delta t, 
\label{xvdescriptiondelta1} \\
v_i(t+\Delta t) & = & v_i(t) - \beta v_i(t) \Delta t 
+ 
\beta \sqrt{2 D \Delta t} \, \zeta_i,
\label{xvdescriptiondelta2}
\end{eqnarray}
where $\beta$ is the (large) friction coefficient and 
$\zeta_i$ is a normally 
distributed random variable with zero mean and unit variance.
This random walk is essentially the Euler scheme for the stochastic 
Langevin equation \cite{Chandrasekhar:1943:SPP}. 
The partially reflective boundary condition at $x=0$ can be stated
as follows: {\it whenever a molecule hits the boundary, 
it is adsorbed with probability $P_4/\sqrt{\beta}$, and reflected otherwise}.
Here, $P_4$ is a given nonnegative number. The implementation
of this boundary condition is straightforward. If $x_i(t+\Delta t)$
computed by (\ref{xvdescriptiondelta1}) is negative then
\begin{equation}
\fl\left.
\begin{array}{ccc}
x_i(t + \Delta t) & = & - x_i(t) - v_i(t) \Delta t  \\
v_i(t + \Delta t) & = & - v_i(t) + \beta v_i(t) \Delta t 
- \beta \sqrt{2 D \Delta t} \, \zeta_i
\end{array}
\; \right\}
\quad \mbox{with probability} \; 1 - \frac{P_4}{\sqrt{\beta}}\,;
\label{partadsmod4}
\end{equation}
otherwise, we remove the $i$-th molecule from the system.
It can be shown that this velocity jump process leads to diffusion 
equation (\ref{diffusionequation1D}) provided that $\beta$ is 
sufficiently large (see Section \ref{secmod4math} for details). 
The parameter $K$ of Robin boundary condition (\ref{RobinBD}) 
is
\begin{equation}
K
=
\frac{P_4 \sqrt{D}}{\sqrt{2 \pi}},
\quad
\mbox{which is equivalent to}
\quad
P_4
=
\frac{K \sqrt{2 \pi}}{\sqrt{D}}.
\label{Kmod4}
\end{equation}

\section{Comparison of stochastic models of diffusion}

\label{illcomp}

In this section, we present the results of two illustrative 
numerical simulations. First, we choose the macroscopic value
of $K$ in (\ref{RobinBD}), and we show the results of
stochastic simulations with the correct choice of 
the probabilities of the adsorption on the reactive boundary
for each model. We demonstrate numerically the validity
of relations between these probabilities and $K$, which
were stated in Section \ref{sec4models} (the mathematical
justification of these formulas is provided later, 
in Section \ref{secmath}). In the second  numerical 
example, we choose the apparently same microscopic
boundary condition for each model. The goal is to
demonstrate that the realistic boundary condition has to be
chosen for each model with care, by applying 
formulas $(\ref{Kmod1})$, $(\ref{Kmod2})$, $(\ref{Kmod3})$ 
or $(\ref{Kmod4})$.

\subsection{Stochastic simulation of models of diffusion}

\label{secil1}

Let us consider the computational domain $[0,5]$, i.e. $L=5$ 
in this section. We choose diffusion constant $D=1$, and reactivity
of the boundary $x=0$ as $K=2$. We consider that the
right boundary $x=L$ is reflecting. Given an initial density
profile $n(x,0)$, we can compute the density $n(x,t)$ 
at time $t \ge 0$ by solving diffusion equation (\ref{diffusionequation1D}) 
accompanied with Robin boundary condition (\ref{RobinBD})
at $x=0$ and no-flux boundary condition at $x=L$.
In this section, we show that comparable results can be obtained
by all four stochastic models provided that we choose
the boundary conditions accordingly.

\begin{figure}
\centerline{
\;\;\;
\psfig{file=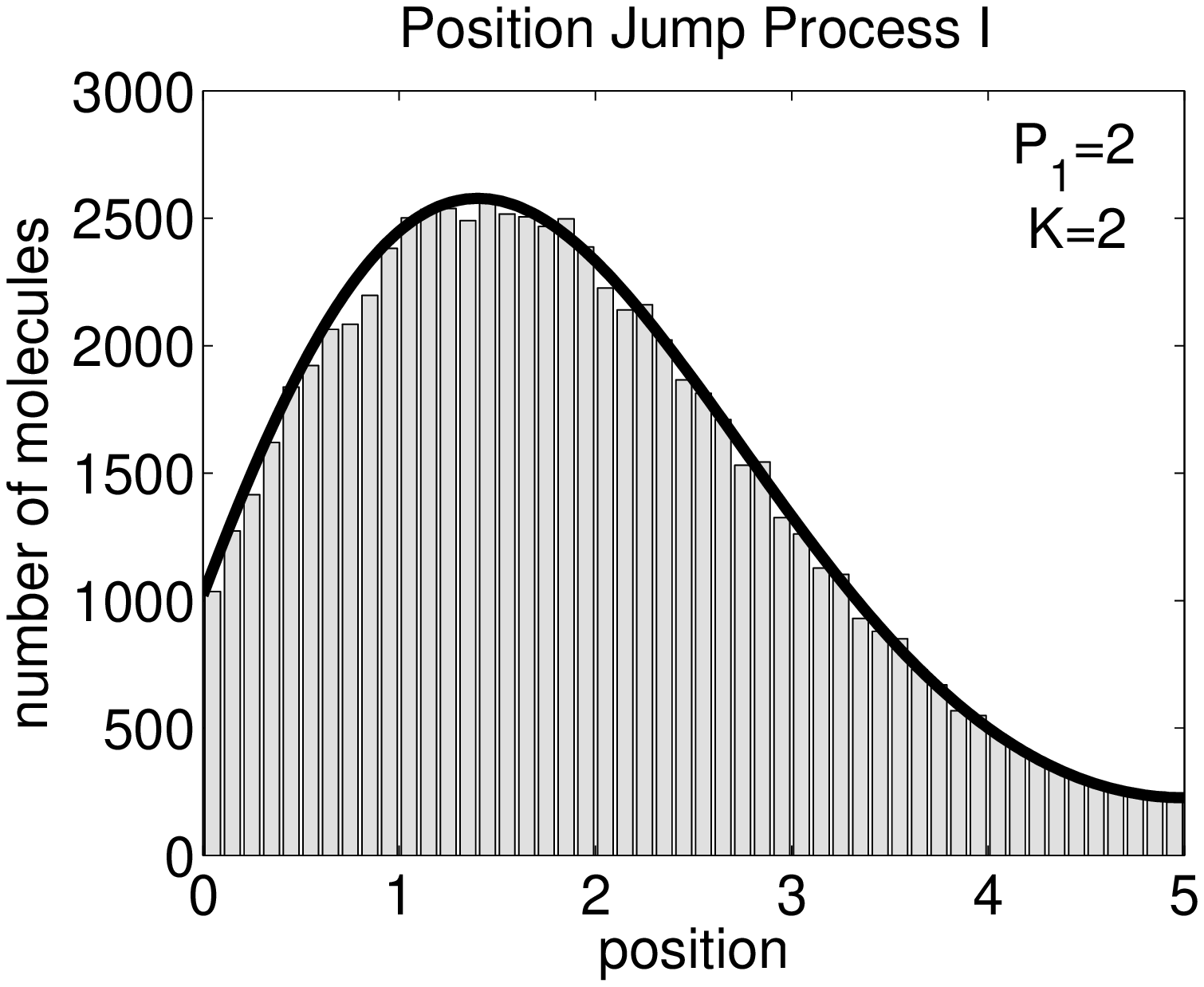,height=2.45in}
$\!\!\!\!\!$
\psfig{file=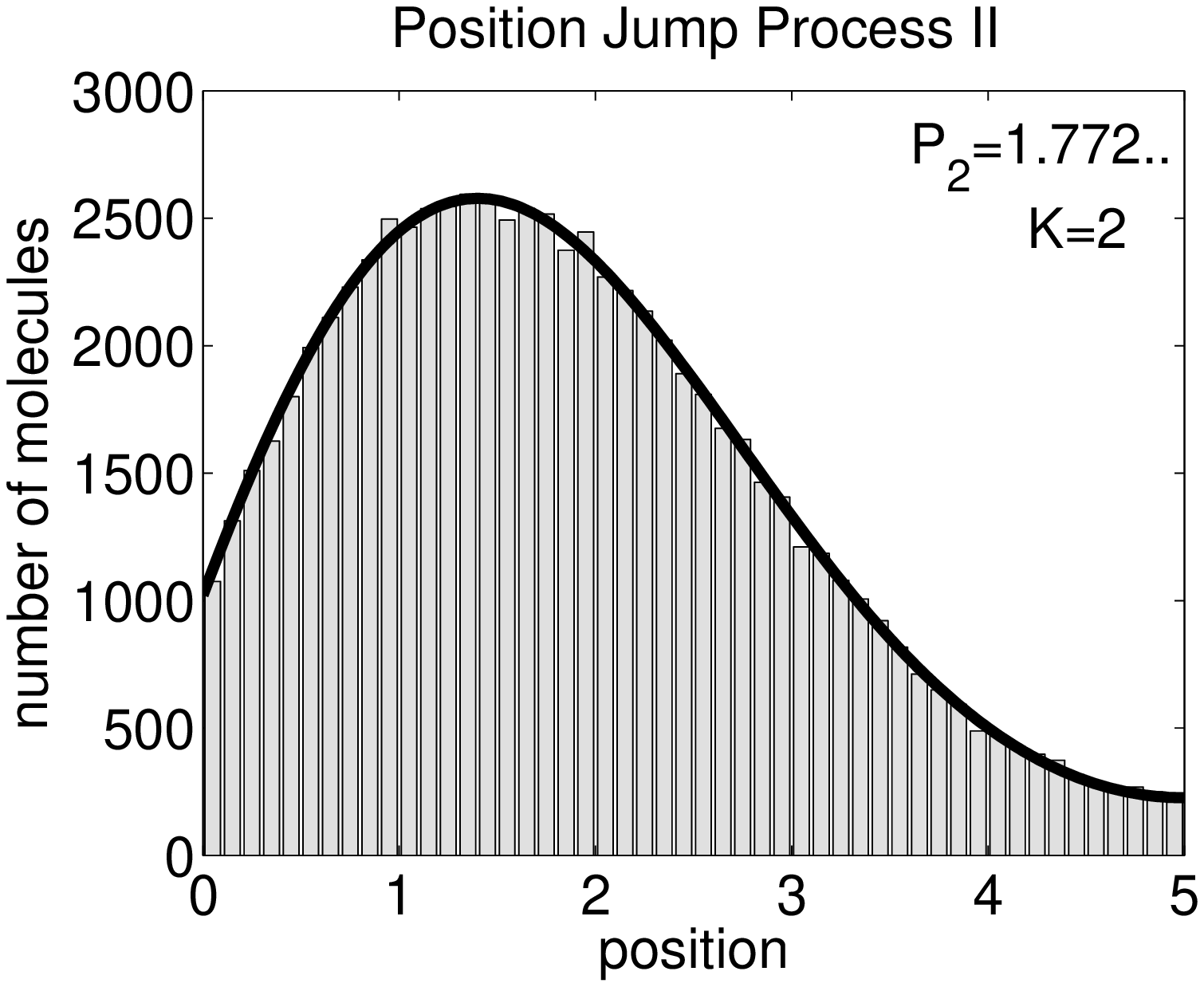,height=2.45in}
}
\bigskip
\centerline{
\;\;\;
\psfig{file=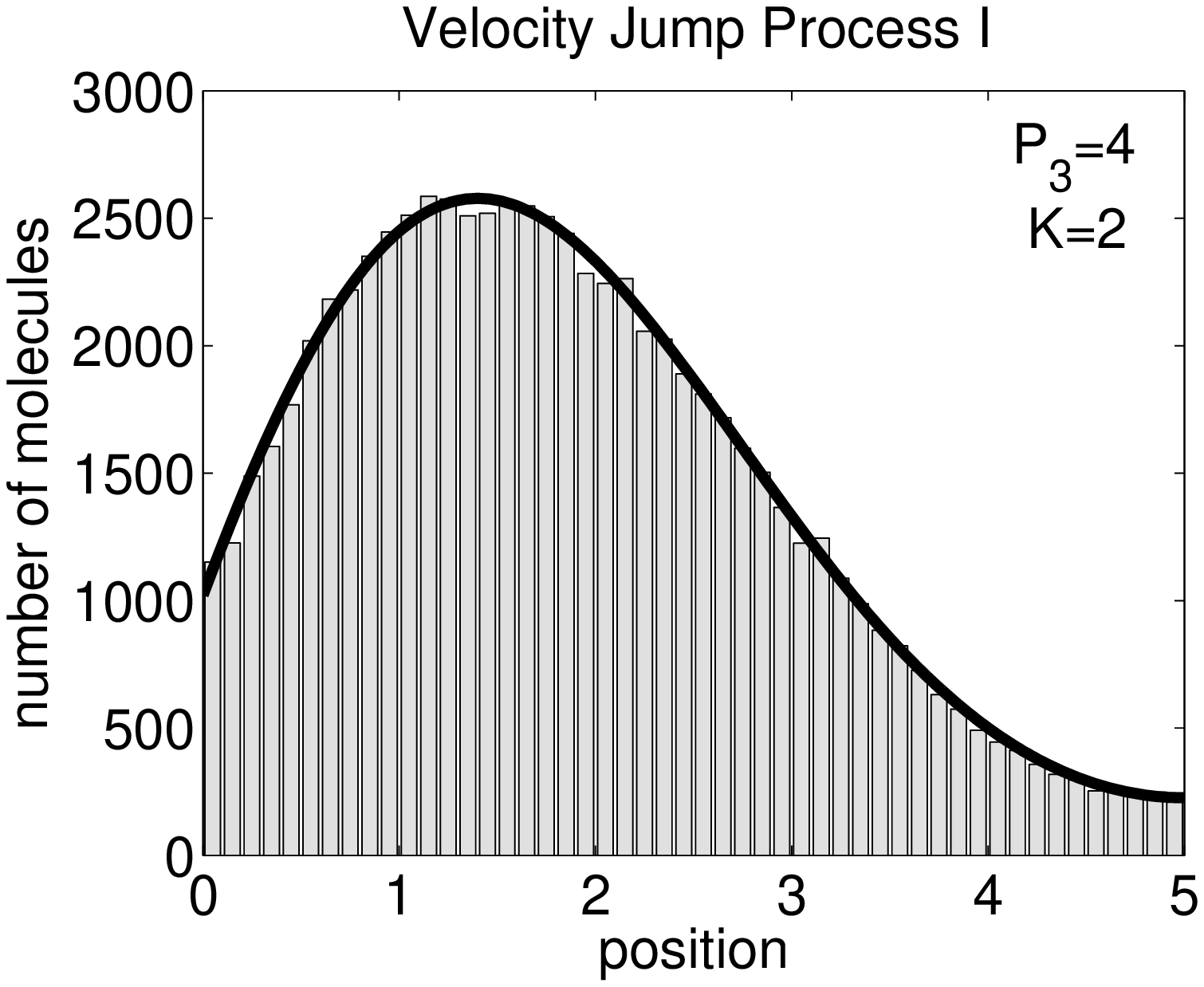,height=2.45in}
$\!\!\!\!\!$
\psfig{file=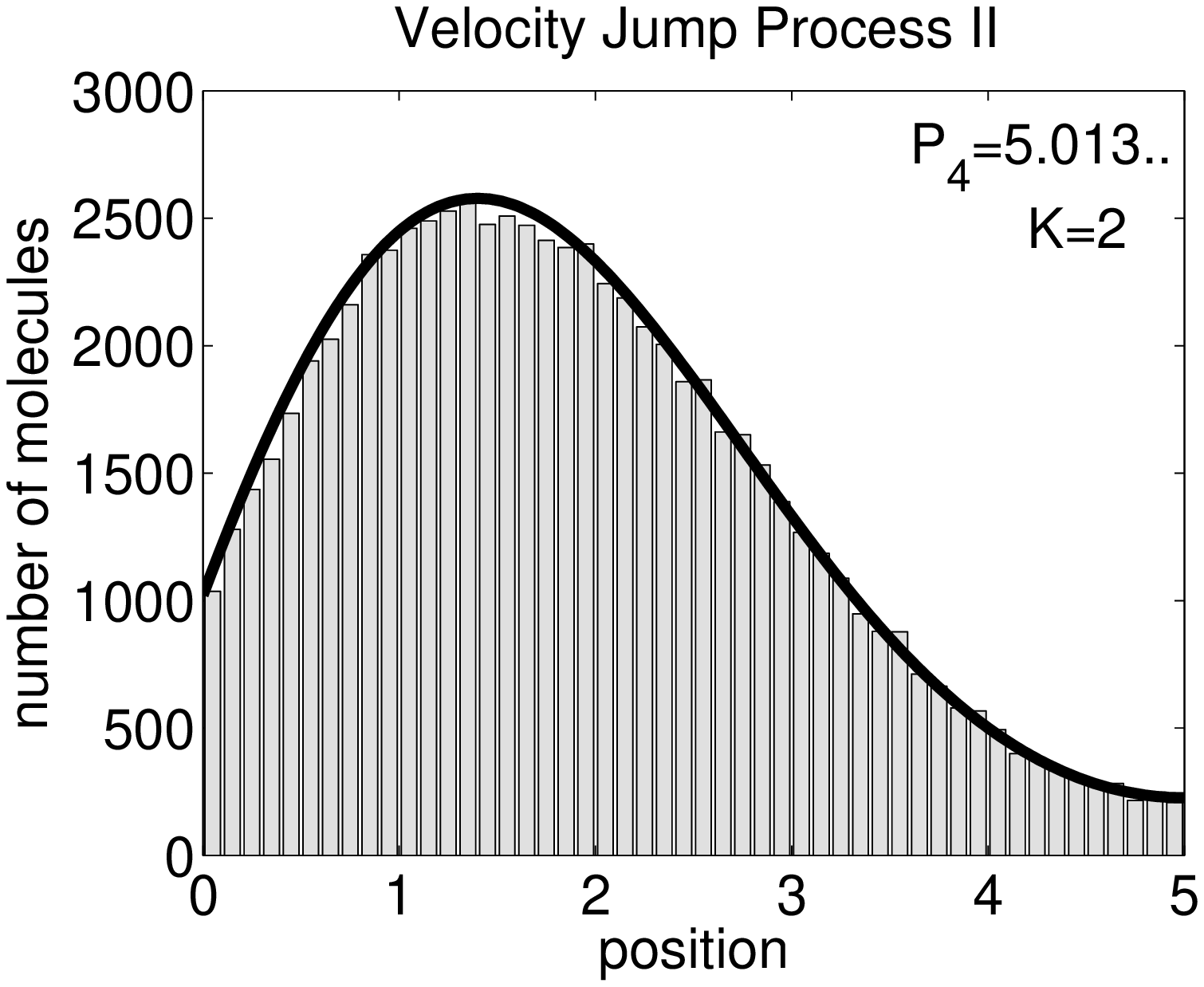,height=2.45in}
}
\caption{{\it Stochastic simulations of four
different diffusion models for $K=2$ and $D=1$. 
Probabilities of adsorption at partially adsorbing 
boundary $x=0$ were computed for each model according
to formulas $(\ref{Kmod1})$, $(\ref{Kmod2})$, $(\ref{Kmod3})$ 
and $(\ref{Kmod4})$.}} 
\label{figillustrativediffusion1}
\end{figure}

The key formulas were provided in the previous section.
Given values of $K$ and $D$, we can compute the adsorbing
probabilities on the reactive boundary $x=0$ by
formula (\ref{Kmod1}) for Position Jump Process I,
formula (\ref{Kmod2}) for Position Jump Process II,
formula (\ref{Kmod3}) for Velocity Jump Process I
and formula (\ref{Kmod4}) for Velocity Jump Process II.
We obtain appropriate values of constants $P_1$, $P_2$, $P_3$ and $P_4$
which are used in the corresponding stochastic model.
In our case $K=2$ and $D=1$, so that
formulas (\ref{Kmod1}), (\ref{Kmod2}), (\ref{Kmod3}) 
and (\ref{Kmod4}) imply
\begin{equation}
P_1 = 2,
\quad
P_2 = \sqrt{\pi} \doteq 1.772,
\quad
P_3 = 4,
\quad
P_4 = 2 \sqrt{2 \pi} \doteq 5.013.
\label{valuesofPK2}
\end{equation}
The results of stochastic simulations are presented in Figure
\ref{figillustrativediffusion1}.
The initial condition was chosen as follows: we start
with $100,000$ molecules in domain $[0,5]$.
We put $75,000$ molecules to position $x=1$ and
$25,000$ to position $x=2$ at time $t=0$. We plot
the density profile at time $t=1$ in Figure
\ref{figillustrativediffusion1}. To do that,
we divide the interval $[0,5]$ into 50 bins of length
$0.1$ and we plot the number of molecules in each
bin at time $t=1$ (histograms). We also plot the solution
of diffusion equation (\ref{diffusionequation1D}) 
accompanied by Robin boundary condition (\ref{RobinBD})
at $x=0$ and no-flux boundary condition at $x=L$.
We see that all four models of diffusion
give the same results provided that we choose the
partially adsorbing probability accordingly.

To compute the simulation results from Figure
\ref{figillustrativediffusion1}, we also had to specify
the additional parameters of the stochastic models. We used
space step $h=0.1$ and time step $\Delta t= 10^{-4}$
in Position Jump Process I. We used time step
$\Delta t = 10^{-4}$ in Position Jump Process II.
We used speed $s=40$ and time step $\Delta t=10^{-5}$
in Velocity Jump Process I and we used friction
coefficient $\beta = 200$ and time step 
$\Delta t=10^{-6}$ in Velocity Jump Process II. 

\subsection{Consequences of the same probability of adsorption}

\begin{figure}
\centerline{
\;\;\;
\psfig{file=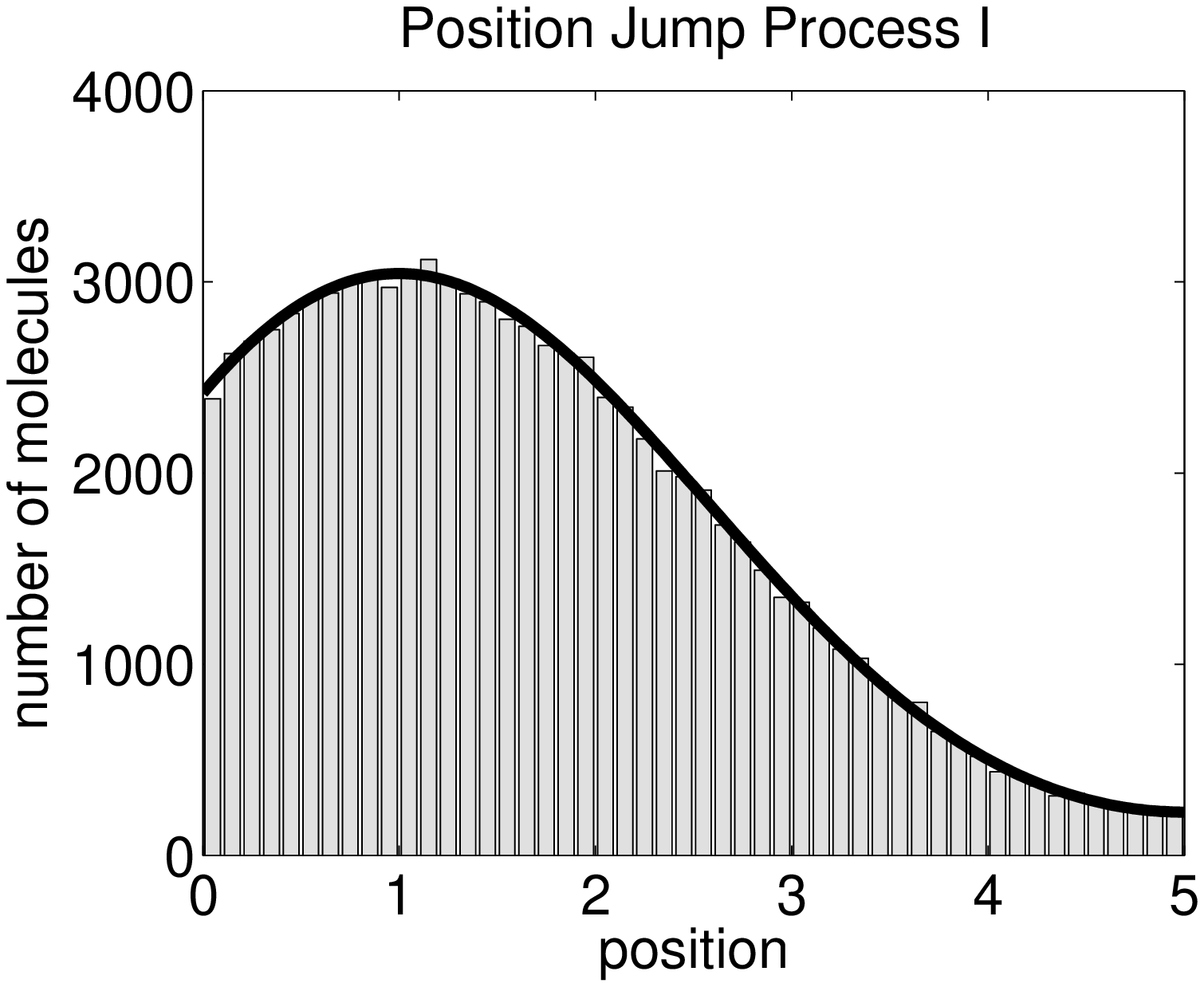,height=2.45in}
$\!\!\!\!\!$
\psfig{file=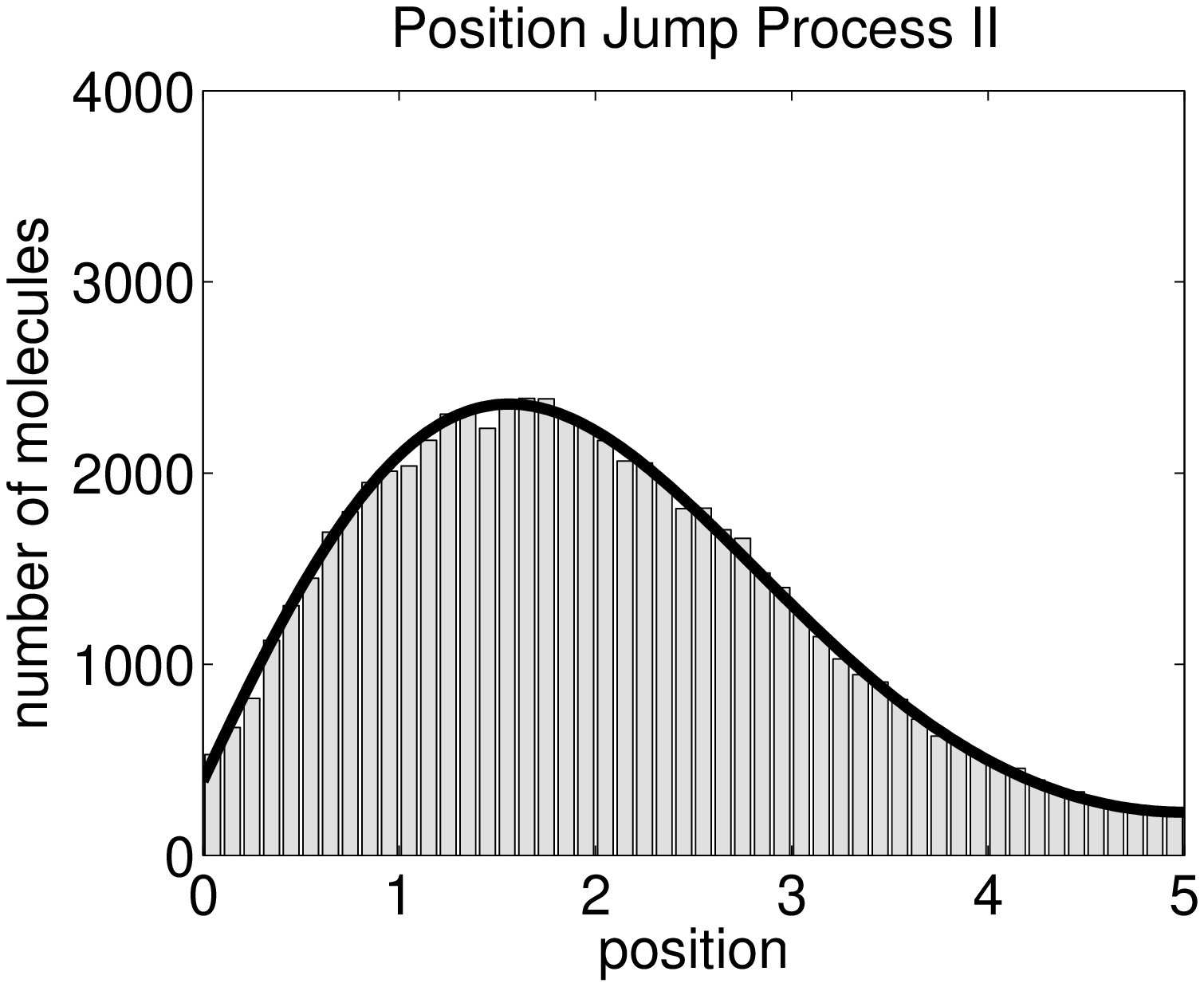,height=2.45in}
}
\bigskip
\centerline{
\;\;\;
\psfig{file=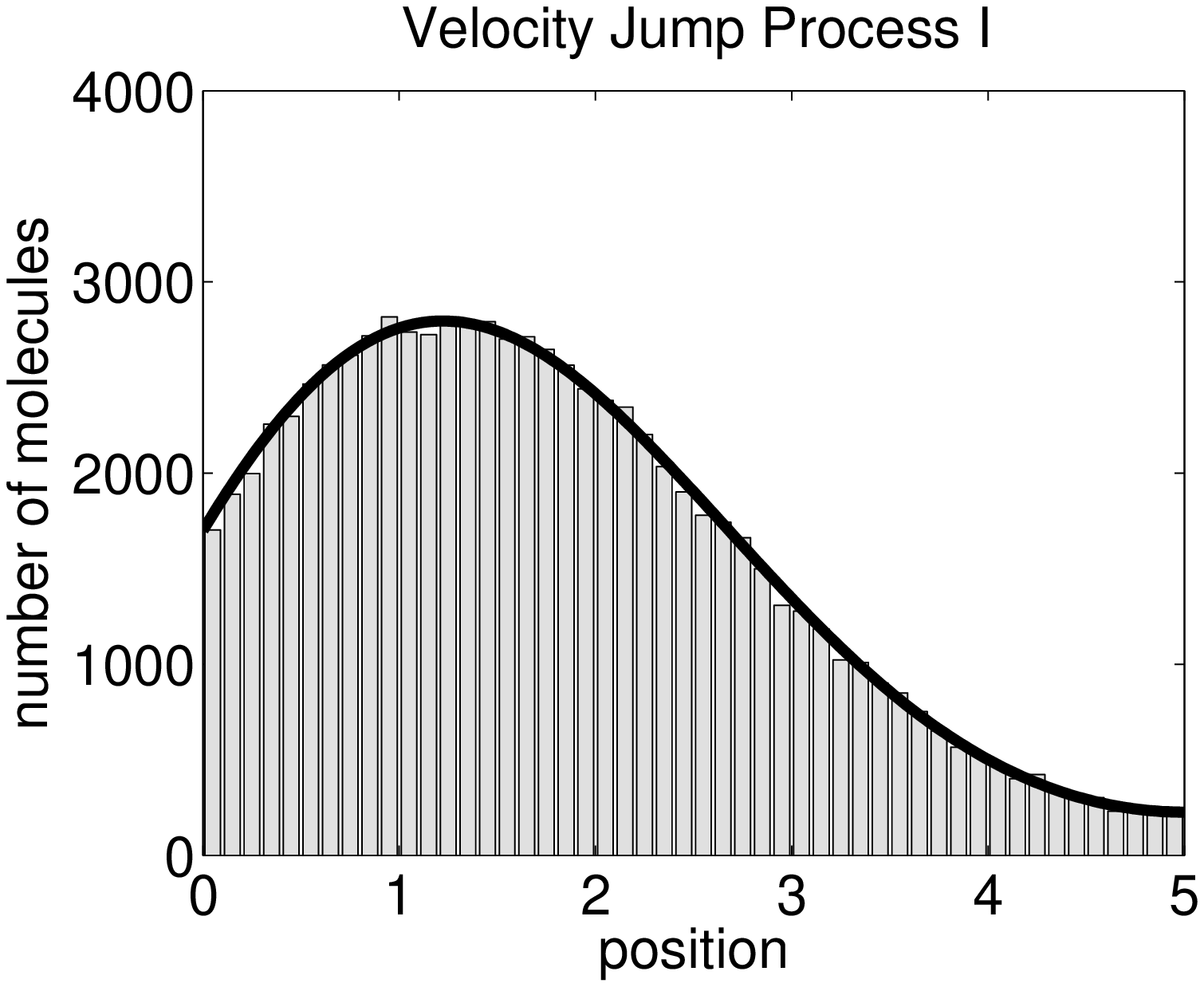,height=2.45in}
$\!\!\!\!\!$
\psfig{file=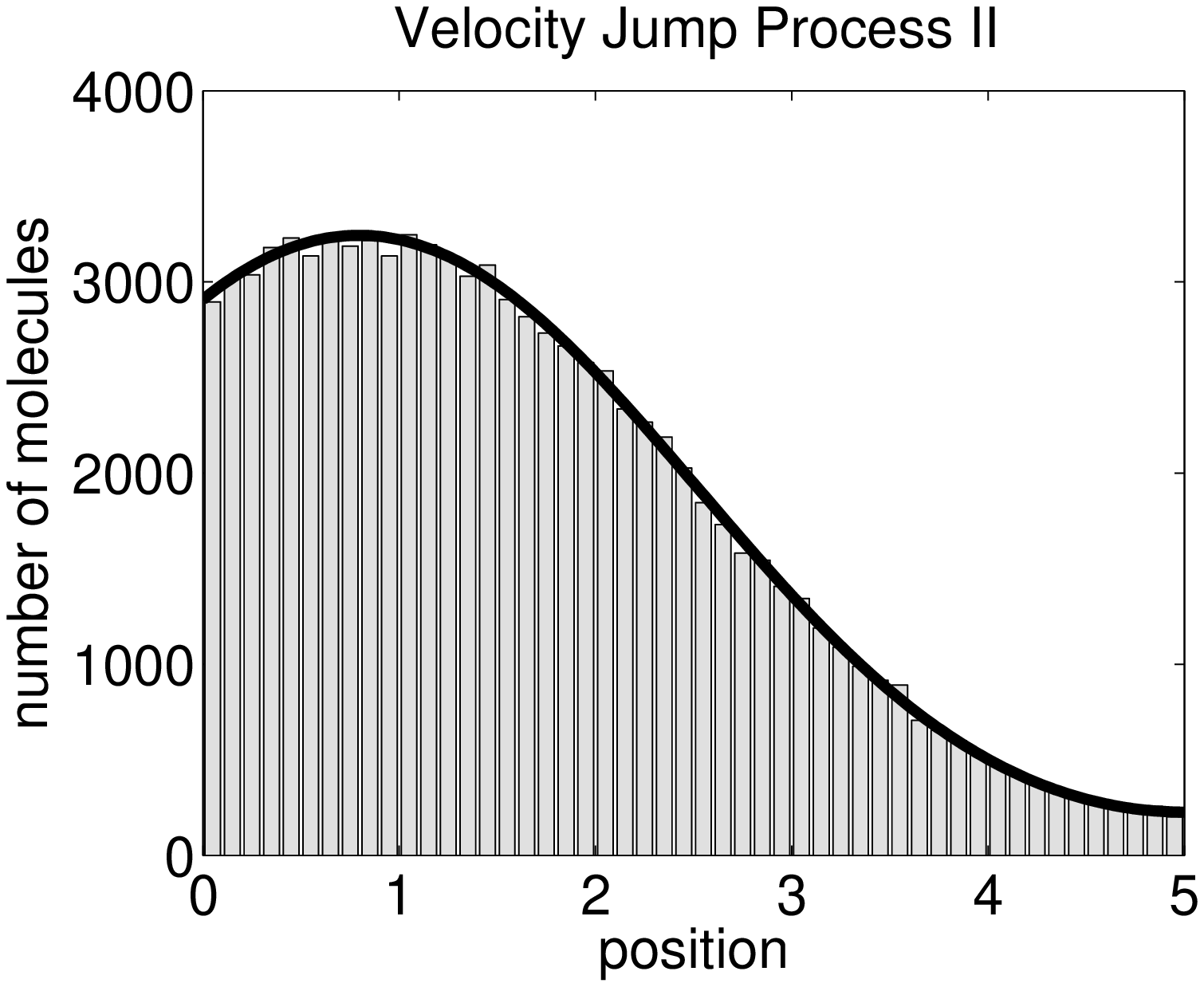,height=2.45in}
}
\caption{{\it Stochastic simulations of four
different diffusion models for $R=0.05$ and $D=1$
(histograms). Solid curves show 
the solution of the diffusion equation $(\ref{diffusionequation1D})$ 
accompanied by no-flux boundary condition at $x=5$
and Robin boundary condition $(\ref{RobinBD})$ at $x=0$ 
where the values of $K$ are computed according
to formulas $(\ref{Kmod1})$, $(\ref{Kmod2})$, $(\ref{Kmod3})$ 
and $(\ref{Kmod4})$.}} 
\label{figillustrativediffusion2}
\end{figure}

Let us now consider the following boundary condition: {\it whenever a molecule 
hits the boundary, it is adsorbed with probability $R$, and reflected
otherwise}. We can easily modify the computer
codes which were used to compute stochastic simulation
results from Figure \ref{figillustrativediffusion1} to incorporate
this boundary condition: whenever
the $i$-th molecule hits the boundary, we generate a random
number $r_i$ uniformly distributed in $[0,1]$. If $r_i < R$,
we remove the $i$-th molecule from the system. It means that the adsorbing
condition which was stated in terms of $P_1$, $P_2$, $P_3$ and $P_4$
is replaced by the same condition expressed in terms of $R$.
Choosing $R=0.05$ and the same parameters and initial
condition as in Section \ref{secil1}, 
the stochastic simulation results at time $t=1$ are shown in Figure 
\ref{figillustrativediffusion2} (histograms obtained
by dividing domain $[0,5]$ into 50 bins and plotting
the number of molecules in each bin).
We clearly see that the results quantitatively differ.
The reason is that the probability of adsorption scales
with other parameters of simulations: namely 
with space step $h$ for Position Jump Process I, 
with time step $\Delta t$ for Position Jump Process II, 
with speed $s$ for Velocity Jump Process I
and with friction coefficient $\beta$ with
Velocity Jump Process II. The values of these
scaling parameters were chosen as in Section 
\ref{secil1}. Namely, we used space step $h=0.1$ 
in Position Jump Process I, time step
$\Delta t = 10^{-4}$ in Position Jump Process II,
speed $s=40$ in Velocity Jump Process I and 
friction coefficient $\beta = 200$ in 
Velocity Jump Process II. Using these values,
we can compute $P_1$, $P_2$, $P_3$ and $P_4$
which correspond to $R=0.05$. Moreover, we can
use formulas $(\ref{Kmod1})$, $(\ref{Kmod2})$, $(\ref{Kmod3})$ 
and $(\ref{Kmod4})$ to compute the corresponding
value of $K$ in Robin boundary condition 
(\ref{RobinBD}). We obtain
$K=0.5$ for Position Jump Process I,
$K\doteq 5.64$ for Position Jump Process II,
$K=1$ for Velocity Jump Process I and
$K\doteq 0.28$ for Velocity Jump Process II.
The solutions of diffusion equation (\ref{diffusionequation1D}) 
accompanied by no-flux boundary condition at $x=5$
and Robin boundary condition (\ref{RobinBD}) at $x=0$
with the appropriate choice of $K$ are plotted
in Figure \ref{figillustrativediffusion2} for comparison
as solid curves. We see that the Robin boundary condition 
(\ref{RobinBD}) at $x=0$ with the appropriate choice of $K$
gives the correct results when compared with stochastic simulations.
Moreover, we also confirm that the same value of $R$
leads to the different values of $K$. Hence the boundary
condition cannot be formulated in terms of one probability
$R$. It has to be appropriately scaled as shown in 
Section \ref{sec4models}. 

To enable a direct comparison between models, we can
slightly reformulate Position Jump Process I. The 
formulation from Section \ref{secmod1des} was chosen
in a way which is used in the stochastic reaction-diffusion
approaches which are based on the reaction-diffusion master 
equation \cite{Hattne:2005:SRD, Isaacson:2006:IDC}.
In particular, no relation between $h$ and $\Delta t$
was given and the probability of partial adsorption
had to be scaled with $h$. One can also formulate
the position jump process on lattice as follows:
we choose time step $\Delta t$ and space step
$h = \sqrt{2 D \, \Delta t}$. At each time step,
the molecule jumps to the left with probability
$1/2$ and to the right otherwise. This random walk
can be accompanied with partially
adsorbing boundary condition: 
{\it whenever a molecule hits the boundary, it is 
adsorbed with probability $\widetilde{P}_1 \sqrt{\Delta t}$, 
and reflected otherwise}. This boundary
condition leads to Robin boundary condition (\ref{RobinBD})
with $K$ given by
\begin{equation*}
K
= \frac{\widetilde{P}_1 \sqrt{D}}{\sqrt{2}}.
\end{equation*}
Comparing this formula with (\ref{Kmod2})
or (\ref{Kmod3}), we see that the Robin boundary
condition is different for Position Jump Process I
and Position Jump Process II even if we scale the
adsorption probability with the same factor
$\sqrt{\Delta t}$. 

The velocity jump processes can also be further compared.
To do this, let us note that the speed $s$ of a molecule
can be estimated as $\sqrt{kT/m}$ where $k$ is the Boltzmann's 
constant, $T$ is the absolute temperature and $m$ is 
the mass of a molecule. In particular, we get the
relation $s = \sqrt{D \beta}$ which can be used to
scale the boundary condition of Velocity Jump Process I
in terms of $\sqrt{\beta}$ instead of $s$. Consequently,
we can relate $P_3$ to $P_4$ by $P_3 = P_4 \sqrt{D}$.
However, using this relation in (\ref{Kmod3}), we obtain
a different Robin boundary condition (\ref{RobinBD})
than by using formula (\ref{Kmod4}). 

To summarize this section, it is possible to reformulate
Position Jump Process I to have the adsorption
probability of both position jump processes scaled
as $P \sqrt{\Delta t}$. It is possible to relate $s$
to $\beta$ in Velocity Jump Process I to have
the adsorption probability of both velocity jump processes 
scaled as $P / \sqrt{\beta}$. Then all four cases 
lead to Robin boundary condition (\ref{RobinBD}) of the 
form 
\begin{equation}
K
=
\alpha P \sqrt{D}
\label{Kmodgenform}
\end{equation}
where $\alpha$ is model-dependent. Consequently,
the boundary condition has to be chosen differently
for each model to get the same value of $K$
in Robin boundary condition (\ref{RobinBD}).
One has to use formulas $(\ref{Kmod1})$, $(\ref{Kmod2})$, $(\ref{Kmod3})$ 
and $(\ref{Kmod4})$ as we showed in Section \ref{secil1}.

\section{Mathematical analysis of stochastic models of diffusion}

\label{secmath}

The goal of this section is to provide the justification
of the results from Section \ref{sec4models}. For each stochastic model, 
we show that the model leads to the diffusion equation 
(\ref{diffusionequation1D}) away from the boundary. Moreover, 
we derive the Robin boundary condition for each model.

\subsection{Position Jump Process I}

\label{secmod1math}

Let $p_k(t)$ be the probability of finding a molecule at mesh point 
$x_k=(2k-1)h/2$ where $k = 1, 2, \dots, M$. If $k \ne 1$, $k \ne M$,
then $p_k$ satisfies
$$
p_k(t + \Delta t)
=
\left(1 - \frac{2 D \Delta t}{h^2} \right) p_k(t)
+
\frac{D \Delta t}{h^2}
\, \Big(
p_{k+1}(t)
+
p_{k-1}(t)
\Big)
$$
which can be rewritten as
$$
\frac{p_k(t + \Delta t) - p_k(t)}{\Delta t}
=
D
\; \frac{
p_{k+1}(t)
+
p_{k-1}(t)
- 
2 p_k(t)
}{h^2}.
$$
Passing to the limit $\Delta t \to 0$, $h \to 0$,
we obtain the diffusion equation (\ref{diffusionequation1D}).
The boundary condition at $x=0$ can be incorporated
into the equation for $p_1(t)$ as
\begin{equation}
p_1(t + \Delta t)
=
\left(1 - \frac{2 D \Delta t}{h^2} \right) p_1(t)
+
\frac{D \Delta t}{h^2}
\, \Big(
p_{2}(t)
+
(1 - P_1 h) 
p_{1}(t)
\Big)
\label{p1equation}
\end{equation}
which can be rewritten as 
$$
\sqrt{\Delta t} \,\,
\frac{p_1(t + \Delta t) - p_1(t)}{\Delta t}
=
\frac{D \sqrt{\Delta t}}{h}
\left(
\frac{p_2(t) - p_1(t)}{h}
-
P_1 \, p_1(t)
\right).
$$
Passing to the limit $\Delta t \to 0$, $h \to 0$ such
that $\sqrt{\Delta t}/h$ is kept constant,
we obtain (\ref{Kmod1}).

\subsection{Position Jump Process II}

\label{secmod2math}

Position jump process (\ref{discretestochevol}) is a 
discretized version (Euler scheme) of the stochastic
differential equation
\begin{equation}
X(t + dt) = X(t) + \sqrt{2 D} \, dW(dt) 
\label{SDEmod2}
\end{equation}
where $dW(dt)$ is the normal random 
variable with mean 0 and variance $dt$ (i.e. the propagator of the special 
Wiener process) and $D$ is the macroscopic diffusion constant. 
The diffusion equation (\ref{diffusionequation1D}) is the
Fokker-Planck equation corresponding to the stochastic
differential equation (\ref{SDEmod2}) and its derivation
can be found in any standard textbook \cite{Risken:1989:FPE}.
The derivation of Robin boundary condition (\ref{Kmod2})
is more delicate and requires the application of 
asymptotic methods \cite{Bender:1978:AMM}. 
To do that, we consider the Position
Jump Process II on the semiinfinite interval $[0,\infty)$
subject to the boundary condition (a)-(b) from Section
\ref{secmod2des}. Let $p_{\Delta t} \equiv p_{\Delta t}(x,t): [0, \infty) 
\times \Delta t \, \en_0 \to [0, \infty)$ be the probability
density function of the discretized process (\ref{discretestochevol})
with the boundary condition (a)-(b), so that
$p_{\Delta t}(x, i \Delta t) \dx$ is the probability
of finding a molecule in the interval $[x, x + \dx]$ 
at time $t=i \Delta t.$ We have
\begin{equation}
p_{\Delta t}(x,t + \Delta t)
=
\int_0^\infty
p(x,t+\Delta t\,|\,y,t) \,
p_{\Delta t}(y,t)
\,\dy,
\label{onestepprob}
\end{equation}
where $p(x,t+\Delta t\,|\,y,t)$ is the conditional probability
distribution function of finding a molecule at point $x$ at time 
$t+\Delta t$ given that it is at point $y$ at time $t.$ There 
are two possible options to reach point $x$ at time $t + \Delta t$: 
either we use (\ref{discretestochevol}) only, i.e. 
$x = y + \sqrt{2 D \, \Delta t} \; \zeta_i$; or we use the
boundary condition $x = - y - \sqrt{2 D \, \Delta t} \; \zeta_i$
with probability $1 - P_2 \sqrt{\Delta t}$. If the former
is true, we have to take into account that some molecules
are lost because of the Andrews and Bray boundary correction (b).
Consequently, we have
\begin{eqnarray*}
\fl
p(x,t+\Delta t\,|\,y,t) 
=
\frac{1}{\sqrt{4 \pi D \, \Delta t}}
\left(
\left\{
1 
-
\exp \left[ - \frac{x y}{D \Delta t} \right] P_2 \sqrt{\Delta t} 
\right\}
\exp \left[ - \frac{(x - y)^2}{4 D \, \Delta t} \right]
+ 
\right.
\\ 
\left.
\qquad + \; (1 - P_2 \sqrt{\Delta t})
\exp \left[ - \frac{(x + y)^2}{4 D \, \Delta t} \right]
\right)
\\
=
\frac{1}{\sqrt{4 \pi D \, \Delta t}}
\left(
\exp \left[ - \frac{(x - y)^2}{4 D \, \Delta t} \right]
+
(1 - 2 P_2 \sqrt{\Delta t})
\exp \left[ - \frac{(x + y)^2}{4 D \, \Delta t} \right]
\right).
\end{eqnarray*}
Thus (\ref{onestepprob}) reads as follows
\[
\fl p_{\Delta t}(x,t + \Delta t)
=
\]
\begin{equation}
\fl
\int_0^\infty 
\frac{p_{\Delta t}(y,t)}{\sqrt{4 \pi D \, \Delta t}}
\left(
\exp \left[ - \frac{(x - y)^2}{4 D \, \Delta t} \right]
+
(1 - 2 P_2 \sqrt{\Delta t})
\exp \left[ - \frac{(x + y)^2}{4 D \, \Delta t} \right]
\right)
\dy.
\label{onestepprobMOD2}
\end{equation}
Away from the boundary, a steepest descent approximation
to the integral as $\Delta t \to 0$ leads to the diffusion
equation (\ref{diffusionequation1D}). However, as
observed in \cite{Singer:2006:PRD}, in the vicinity of
the boundary, such a calculation needs to be modified:
there is a boundary layer of width $\sqrt{\Delta t}$,
and it is the solution in the boundary layer which determines
the boundary condition of the diffusion equation.
In the boundary layer, we change variables from $x$ to
$\eta$ by setting $x = \sqrt{\Delta t} \, \eta$ and define the inner
solution
$$
p_{inner} (\eta, t)
=
p_{\Delta t}(\sqrt{\Delta t} \, \eta, t).
$$
Expanding $p_{inner}$ in the powers of $\sqrt{\Delta t}$, we obtain
$$
p_{inner}(\eta, t)
\sim
p_{i,0} (\eta,t)
+
\sqrt{\Delta t} \,
p_{i,1} (\eta,t)
+
\Delta t \,
p_{i,2} (\eta,t)
+
\dots.
$$
Using this expansion in the integral equation (\ref{onestepprobMOD2})
and comparing the terms of the same order, we obtain 
at $O(1)$ that $p_{i,0}$ is independent of $\eta$. 
At $O(\sqrt{\Delta t})$ we find that
$$
p_{i,1}(\eta)
=
-
2 P_2
\int_0^\infty 
\frac{p_{i,0}}{\sqrt{4 \pi D}}
\exp \left[ - \frac{(\xi + \eta)^2}{4 D} \right]
\dxi
$$
\begin{equation}
+
\int_0^\infty 
\frac{p_{i,1}(\xi)}{\sqrt{4 \pi D}}
\left(
\exp \left[ - \frac{(\xi - \eta)^2}{4 D} \right]
+
\exp \left[ - \frac{(\xi + \eta)^2}{4 D} \right]
\right)
\dxi.
\label{onestepprob4}
\end{equation}
Now by matching the inner boundary layer expansion
with the outer expansion $p_{\Delta t} (x,t) \sim n(x,t) + \dots$,
we find $p_{i,0}(t)=n(0,t)$ and
\begin{equation}
\lim_{\eta \to \infty}
\frac{\partial p_{i,1}}{\partial \eta}(\eta,t)
=
\frac{\partial n}{\partial x}(0,t).\label{match}
\end{equation}
Thus to determine the boundary condition on $n$ at $x=0$ we need to determine
the behaviour of $\partial p_{i,1}/\partial \eta$ as $\eta \to \infty.$
Differentiating (\ref{onestepprob4}) with respect to $\eta$, we obtain
$$
\fl \frac{\partial p_{i,1}}{\partial \eta}(\eta)
=
\frac{P_2 \, p_{i,0}}{\sqrt{\pi D}} 
\exp 
\left[ 
- \frac{\eta^2}{4D} 
\right]
$$
$$
+
\int_0^\infty 
\frac{p_{i,1}(\xi)}{\sqrt{4 \pi D}}
\left(
\frac{\xi - \eta}{2D}
\exp \left[ - \frac{(\xi - \eta)^2}{4 D} \right]
-
\frac{\xi + \eta}{2D}
\exp \left[ - \frac{(\xi + \eta)^2}{4 D} \right]
\right)
\dxi.
$$
Using integration by parts, we obtain the integral equation
\begin{equation}
\fl\frac{\partial p_{i,1}}{\partial \eta}(\eta)
=
\frac{P_2 \, p_{i,0}}{\sqrt{\pi D}} 
\exp 
\left[ 
- \frac{\eta^2}{4D} 
\right]
\label{integraleq}
\end{equation}
$$
+\frac{1}{\sqrt{4 \pi D}}
\int_0^\infty 
\frac{\partial p_{i,1}}{\partial \eta}(\xi)
\left(
\exp \left[ - \frac{(\xi - \eta)^2}{4 D} \right]
-
\exp \left[ - \frac{(\xi + \eta)^2}{4 D} \right]
\right)
\dxi.
$$
Let us define the function $g(\eta)$ by
\begin{equation}
g(\eta)
=
-\frac{P_2 \, p_{i,0}}{\sqrt{\pi D}} 
\exp 
\left[ 
- \frac{\eta^2}{4D} 
\right]
+
\frac{\partial p_{i,1}}{\partial \eta}(\eta).
\label{defdeta}
\end{equation}
Then (\ref{integraleq}) can be rewritten as 
\begin{equation}
\fl
g(\eta)
=
\phi(\eta)
+
\frac{1}{\sqrt{4 \pi D}}
\int_0^\infty 
g(\xi)
\left(
\exp \left[ - \frac{(\xi - \eta)^2}{4 D} \right]
-
\exp \left[ - \frac{(\xi + \eta)^2}{4 D} \right]
\right)
\dxi
\label{integraleq3}
\end{equation}
where
\begin{equation}
\phi(\eta)
=
\frac{P_2 \, p_{i,0}}{\sqrt{8 \pi D}}
\, \exp \left[ - \frac{\eta^2}{8 D} \right]
\left(
\erf \left[ \frac{\eta}{\sqrt{8 D}} \right]
-
\erf \left[ - \frac{\eta}{\sqrt{8 D}} \right]
\right)
\label{defphi}
\end{equation}
and the error function is defined by
\begin{equation*}
\erf(\xi) = \frac{2}{\sqrt{\pi}}
\int_0^\xi \exp \left[ - \sigma^2 \right] \dsigma.
\end{equation*}
The function $g(\eta)$ is defined for $\eta \ge 0$.
Since $\phi(\eta)$ is odd function, we can define
$g(\eta)$ for the negative values as an odd function too by setting
 $g(\eta) = - g(-\eta)$ for $\eta < 0$. Then
equation (\ref{integraleq3}) can be simplified to
\begin{equation}
g(\eta)
=
\phi(\eta)
+
\frac{1}{\sqrt{4 \pi D}}
\int_{-\infty}^\infty 
g(\xi)
\exp \left[ - \frac{(\xi - \eta)^2}{4 D} \right]
\dxi.
\label{integraleq4}
\end{equation}
The natural way to solve such an equation is to apply a Fourier
transform, but we have to be slightly careful since the Fourier 
transform of $g$
does not exist in the classical sense ($g$ tends to a constant at
infinity, so is not integrable).
Defining 
\[ 
g_+(\eta) = g(\eta) \chi_{[0,\infty)}(\eta)\qquad
g_-(\eta) = g(\eta) \chi_{(-\infty,0]}(\eta),
\]
where $\chi_{[a,b]}$ is the characteristic function of the interval
$[a,b]$ (that is, $\chi_{[a,b]}(\eta) = 1$ if $a\leq \eta\leq b$ and
zero otherwise), and applying the Fourier transform 
\[
\widehat{h}(k) = \int_{-\infty}^\infty h(\eta)
\exp[ i k \eta] \, \deta
\]
to equation (\ref{integraleq4}), we obtain
\begin{equation*}
\widehat{g_+}(k)+\widehat{g_-}(k)
=
\widehat{\phi}(k)
+
\left(\widehat{g_+}(k)+\widehat{g_-}(k)\right)
\exp \left[ - D k^2 \right].
\end{equation*}
Thus
\begin{equation}
\widehat{g_+}(k)+\widehat{g_-}(k)
=
\frac{
\widehat{\phi}(k)
}{1-\exp \left[ - D  k^2 \right]}.
\label{formgeta}
\end{equation}
This seems like one equation for the two unknowns $\widehat{g_+}(k)$
and $\widehat{g_-}(k)$, but in fact we know from their definitions
that $\widehat{g_+}(k)$ is analytic for the imaginary part of $k$
positive, while $\widehat{g_-}(k)$ is analytic for the imaginary part
of $k$ negative, and this tells us how to divide all the poles of the
right-hand side between $\widehat{g_+}(k)$
and $\widehat{g_-}(k)$, except for the pole at the origin, which may
appear in either $\widehat{g_+}(k)$
or $\widehat{g_-}(k)$. To divide this pole contribution up we note
that since $g$ is odd, $\widehat{g_+}(k) = -\widehat{g_-}(-k)$,
which implies that the coefficients of the odd powers of $k$ near zero
are equal in  $\widehat{g_+}(k)$ and $\widehat{g_-}(k)$, and that
the coefficients of the even powers are zero.

Using (\ref{defdeta}) we have
\begin{equation*}
\lim_{\eta \to \infty}
\frac{\partial p_{i,1}}{\partial \eta}(\eta)
=
\lim_{\eta \to \infty}
g(\eta)
=
\lim_{\eta \to \infty}
\frac{1}{2 \pi}\int_{-\infty}^\infty
\widehat{g_+}(k) \exp[-ik\eta]
\, \dk,
\end{equation*}
where the inversion contour lies in the upper half-plane.
Deforming the contour to $-i\infty$ we pick up residue contributions
from each of the poles of (\ref{formgeta}) in the lower half
plane. The only finite contribution as $\eta \rightarrow \infty$ arises from
the pole at the origin. Since
\[ \widehat{\phi}(k) \sim 
\frac{ 4i  \sqrt{D} P_2 p_{i,0} k}{\sqrt{\pi}}\mbox{ as }k
\rightarrow 0, 
\]
 we have
\[  \widehat{g_+}(k)\sim \frac{ 2 i    P_2
  p_{i,0}}{\sqrt{\pi D}\,k} 
\mbox{ as }k
\rightarrow 0,\]
so that 
\begin{equation*}
D \lim_{\eta \to \infty}
\frac{\partial p_{i,1}}{\partial \eta}(\eta)
=
\frac{2 P_2 \, \sqrt{D}}{\sqrt{\pi}}\, p_{i,0}.
\end{equation*}
Using the matching condition (\ref{match}) we therefore derive the
Robin boundary 
condition (\ref{Kmod2}) 
for the stochastic boundary condition (a)-(b). If we consider
the boundary condition (a) only, equation (\ref{onestepprob}) leads 
to the modified formula (\ref{onestepprobMOD2}) where $2 P_2$ is 
replaced by $P_2$. Thus the boundary layer 
method presented above gives in that case the 
Robin boundary condition (\ref{Kmod2alt}), 
which differs from (\ref{Kmod2}) by the factor of two.

\subsection{Velocity Jump Process I}

\label{secmod3math}

Using standard methods \cite{Kac:1974:SMR,Erban:2004:ICB}, one can
show that the density of molecules $n(x,t)$ satisfies the damped
wave (telegrapher's) equation
\begin{equation}
\frac{1}{2 \lambda}
\frac{\partial^2 n}{\partial t^2}
+ 
\frac{\partial n}{\partial t}
= 
\frac{s^2}{2 \lambda}
\frac{\partial^2 n}{\partial x^2}.
\label{hypereq}
\end{equation}
The long time behaviour of (\ref{hypereq}) is described by the
corresponding parabolic limit \cite{Zauderer:1983:PDE}. 
Using (\ref{lambdasD}), we obtain that $n(x,t)$ satisfies 
(\ref{diffusionequation1D}) for times $t \gg \lambda^{-1}.$ 

Next, we derive the Robin boundary condition
corresponding to (\ref{partadsveljump}). Let 
$p^+(x,t)$ be the density of molecules that are at $(x,t)$ 
and are moving to the right, and let $p^-(x,t)$ be the density 
of molecules that are at $(x,t)$ and are moving to the left. 
Then the density of molecules at $(x,t)$ is given by
the sum $n(x,t) = p^+(x,t) + p^-(x,t),$ and the 
flux is $j(x,t) = s(p^+(x,t) - p^-(x,t))$. The stochastic
boundary condition (\ref{partadsveljump}) implies
\begin{equation*}
p^+(0,t)
=
\left(
1 - \frac{P_3}{s}
\right) \,
p^-(0,t).
\end{equation*}
This boundary condition can be written in terms of $n$ and $j$ as
\begin{equation*}
\frac{1}{2}
\left(n(0,t) + \frac{j(0,t)}{s} \right)
=
\left(
1 - \frac{P_3}{s}
\right) \,
\frac{1}{2}
\left(n(0,t) - \frac{j(0,t)}{s} \right)
\end{equation*}
which implies
\begin{equation*}
P_3 
\, n(0,t)
=
\left(
\frac{P_3}{s} - 2
\right) \,
j(0,t).
\end{equation*}
Since  
\begin{equation}
j(0,t) \approx - D \frac{\partial n}{\partial x}(0,t), 
\label{jdiflux}
\end{equation}
we derive (\ref{Kmod3}) in the limit $s \to \infty$.

\subsection{Velocity Jump Process II}

\label{secmod4math}

The random walk (\ref{xvdescriptiondelta1})-(\ref{xvdescriptiondelta2})
is a discretized version of Langevin's equation 
\cite{Chandrasekhar:1943:SPP}. To derive the Robin boundary 
condition, we consider the 
behaviour of molecules in the semiinfinite interval $[0,\infty)$.
The $i$-th molecule is described 
by two variables: its position $x_i(t)$ and its velocity $v_i(t)$.
We compute the position $x_i(t+\Delta t)$ and velocity $v_i(t+\Delta t)$ 
from the position $x_i(t)$ and velocity $v_i(t)$
by (\ref{xvdescriptiondelta1})-(\ref{xvdescriptiondelta2})
together with boundary condition (\ref{partadsmod4}) at $x=0$. 
Let $f(x,v,t)$ be the density of molecules which are at position $x$ with 
velocity $v$ at time $t$, so that  
$f(x,v,t) \, \delta x \,\delta v$ is 
number of molecules in interval $[x,x+\delta x]$ with velocities
between $v$ and $v+\delta v$ at time $t$. Assuming that the
change in velocity of the $i$-th molecule during the time step is 
$\Delta v$ (i.e. $\Delta v = v_i(t+\Delta t) -  v_i(t)$), 
there are two possible options 
for the molecule to reach a point $x \ge 0$ with velocity $v$ at time 
$t + \Delta t$: either the molecule was at position 
$x_i(t) = x - (v - \Delta v) \Delta t$ with velocity 
$v_i(t) = v - \Delta v$ at time $t$; or at position
$x_i(t) = - x + (v + \Delta v) \Delta t$ with velocity 
$v_i(t) = - v - \Delta v$ and was reflected according
to (\ref{partadsmod4}). Both cases make sense only
if $x_i(t) \ge 0$. Consequently, $f(x,v,t+\Delta t)$
can be computed from $f(\cdot,\cdot,t)$ by the
integral equation
\begin{equation}
\fl f(x,v,t+\Delta t) 
= 
\int_{v-x/\Delta t}^\infty f(x - (v - \Delta v) \Delta t,v-\Delta v,t)
\, \psi(v -\Delta v; \Delta v) \,\mbox{d}\Delta v
\; +
\label{integralformulafxv}
\end{equation}
$$
+
\left(
1-\frac{P_4}{\sqrt{\beta}}
\right) 
\int_{-v+x/\Delta t}^\infty 
f(-x + (v + \Delta v)\Delta t,-v-\Delta v,t)
\, \psi(-v -\Delta v; \Delta v) \,\mbox{d}\Delta v
$$
where $\psi(w; \Delta v)$ is a distribution function 
of the conditional probability that the change in
velocity during the time step is $\Delta v$ provided that 
$v_i(t) = w.$ Using (\ref{xvdescriptiondelta2}),
we obtain
\begin{equation}
\psi(w; \Delta v)
=
\frac{1}{\beta \sqrt{4 \pi D \Delta t}}
\exp \left[
- \frac{(\Delta v + \beta w \Delta t)^2}{4 \beta^2 D \Delta t}
\right].
\label{psidistribution}
\end{equation}
Passing to the limit $\Delta t \to 0$, we obtain that $f(x,v,t)$ 
satisfies the Fokker-Planck equation \cite{Chandrasekhar:1943:SPP}
\begin{equation}
\frac{\partial f}{\partial t}
+
v \frac{\partial f}{\partial x}
=
\beta
\frac{\partial}{\partial v}
\left(
v f
+
\beta D \frac{\partial f}{\partial v}
\right)
\label{FokkerPlanckvp}
\end{equation}
together with the boundary condition
\begin{equation}
f(0,v,t)
=
\left(
1-\frac{P_4}{\sqrt{\beta}}
\right) 
f(0,-v,t).
\label{boundaryv}
\end{equation}
The density of molecules at the point $x$ and time $t$ is defined by
\begin{equation}
n(x,t) = \int_\er f(x,v,t) \, \dv.
\label{defn}
\end{equation}
To derive the diffusion equation for $n$ and the corresponding 
Robin boundary condition we consider the limit in which $\beta
\rightarrow \infty$
and rescale the velocity variable by setting
$$
v = \eta \sqrt{\beta}, \qquad
\overline{f}(x,\eta,t) = f(x, v, t),
$$
to give
\begin{equation}
\frac{1}{\beta}
\frac{\partial \overline{f}}{\partial t}
+
\frac{1}{\sqrt{\beta}} \, \eta \frac{\partial \overline{f}}{\partial x}
=
\frac{\partial}{\partial \eta}
\left(
\eta \overline{f}
+
D \frac{\partial \overline{f}}{\partial \eta}
\right).
\label{FokkerPlancknup}
\end{equation}
We expand $\overline{f}$ in powers 
of $1/\sqrt{\beta}$ as
\begin{equation}
\overline{f}(x,\eta,t) = f_0(x,\eta,t) 
+ 
\frac{1}{\sqrt{\beta}} \, f_1(x,\eta,t) 
+ \frac{1}{\beta} \, f_2(x,\eta,t) + \dots.
\label{expanoverf}
\end{equation}
Substituting (\ref{expanoverf}) into (\ref{FokkerPlancknup}) and
equating coefficients of powers of $\beta$ we obtain
\begin{eqnarray}
\frac{\partial}{\partial \eta}
\left(
\eta f_0
+
D \frac{\partial f_0}{\partial \eta}
\right)
=
0, 
\label{f0equation}
\\
\frac{\partial}{\partial \eta}
\left(
\eta f_1
+
D \frac{\partial f_1}{\partial \eta}
\right) 
=
\eta \frac{\partial f_0}{\partial x},
\label{f1equation}
\\
\frac{\partial}{\partial \eta}
\left(
\eta f_2
+
D \frac{\partial f_2}{\partial \eta}
\right)
=
\eta \frac{\partial f_1}{\partial x}
+
\frac{\partial f_0}{\partial t}.
\label{f2equation}
\end{eqnarray}
Solving equations (\ref{f0equation})-(\ref{f1equation}), 
we obtain  
\begin{eqnarray}
f_0(x,\eta,t) =  \varrho(x,t) \exp \left[ - \frac{\eta^2}{2 D} \right],
\label{f0sol}
\\
f_1(x,\eta,t) =  - \frac{\partial \varrho}{\partial x}(x,t) 
\, \eta \exp \left[ - \frac{\eta^2}{2 D} \right]
\label{f1sol}
\end{eqnarray}
where the function $\varrho(x,t)$ is independent of $\eta$. Substituting 
(\ref{f0sol})-(\ref{f1sol}) into (\ref{f2equation}) gives
\begin{equation*}
\frac{\partial}{\partial \eta}
\left(
\eta f_2
+
D \frac{\partial f_2}{\partial \eta}
\right)
=
- \frac{\partial^2 \varrho}{\partial x^2} \,
\eta^2 \exp \left[ - \frac{\eta^2}{2 D} \right] 
+
\frac{\partial \varrho}{\partial t} 
\exp \left[ - \frac{\eta^2}{2 D} \right].
\end{equation*}
Integrating over $\eta$ gives the solvability condition
\begin{equation*}
\frac{\partial \varrho}{\partial t} 
=
D \frac{\partial^2 \varrho}{\partial x^2}. 
\end{equation*}
Using (\ref{defn}), we see that $\varrho(x,t)$
is proportional to density of individuals 
$n(x,t)$ for large $\beta$. Consequently, $n(x,t)$ satisfies the 
diffusion equation (\ref{diffusionequation1D}) for large $\beta$.
Multiplying (\ref{boundaryv}) by $v$ and integrating over $v$, we obtain 
\begin{equation}
j(0,t) 
= 
-
\frac{P_4}{\sqrt{\beta}}
\int_0^\infty v f(0,-v,t) \, dv
\label{integralformulafxv2}
\end{equation}
where flux is defined by $j(0,t) = \int_\er v f(0,v,t) \, \dv.$
Substituting (\ref{f0sol})-(\ref{f1sol}) into 
(\ref{integralformulafxv2}), we derive the Robin
boundary condition (\ref{Kmod4}).

\section{Boundary conditions for stochastic models of 
reaction-diffusion processes}

\label{secrd}

In this section, we show that reactions in the solution
do not change the boundary conditions from Section
\ref{sec4models}, i.e. the boundary conditions of stochastic
models of the reaction-diffusion processes are determined
by the corresponding diffusion model. First, we illustrate this 
fact numerically in Sections \ref{secposjump1RD} and
\ref{secposjump2RD}. In Section  \ref{secposjump1RD},
we use the stochastic approach based on the 
reaction-diffusion master equation 
\cite{Hattne:2005:SRD, Isaacson:2006:IDC}, so that 
the corresponding diffusion model is the Position Jump 
Process I. In Section \ref{secposjump2RD}, we use the stochastic 
approach of Andrews and Bray \cite{Andrews:2004:SSC},
so that the corresponding diffusion model is the Position Jump 
Process II. Then, in Section \ref{secmathRD}, we provide mathematical 
justification of the fact that the presence of reactions in the
solution does not influence the
choice of the boundary condition.

\subsection{Nonlinear reaction kinetics}

\label{secposjump1RD}

We consider two chemicals $A$ and $B$ which diffuse
in the domain of interest $[0,1]$ with diffusion
constants $D_A$ and $D_B$, respectively, and 
which react according to  Schnakenberg 
reaction kinetics \cite{Schnakenberg:1979:SCR}. 
The chemical $A$ is produced with a constant
rate (from a suitable reactant which is supposed
to be in excess) and degraded. The chemical $B$
is also produced with a constant rate. Moreover, $A$
and $B$ react according to the cubic reaction
\begin{equation}
2 A + B
\quad
\mathop{\longrightarrow}^{k_c}
\quad
3 A
\label{ABreact}
\end{equation}
where $k_c$ is reaction constant. We use a 
partially adsorbing boundary condition at $x=0$ and a
reflective boundary condition at $x=1$.

The stochastic simulation algorithm is based 
on the reaction-diffusion master 
equation \cite{Hattne:2005:SRD, Isaacson:2006:IDC}.
We divide the domain of interest
into $M$ compartments of the length $h=1/M$ which
are assumed to be well-mixed. In particular,
one can use the classical Gillespie's algorithm
\cite{Gillespie:1977:ESS} to simulate stochastically
the reactions in each compartment. The system is then
described by two $M$-dimensional vectors
$[A_1,A_2, \dots, A_M]$ and $[B_1,B_2, \dots, B_M]$
where $A_i$ (resp. $B_i$) denotes the number of molecules
of chemical $A$ (resp. $B$) in the $i$-th compartment.
The diffusion of chemicals is added to the system as 
another set of reactions--jumps between the neighbouring
compartments with the rate $D_A/h^2$ (resp. $D_B/h^2$)
\cite{Qiao:2006:SDS}. In particular, the model
of diffusion is equivalent to the Position Jump Process I.

\begin{figure}
\centerline{
\;\;\;
\psfig{file=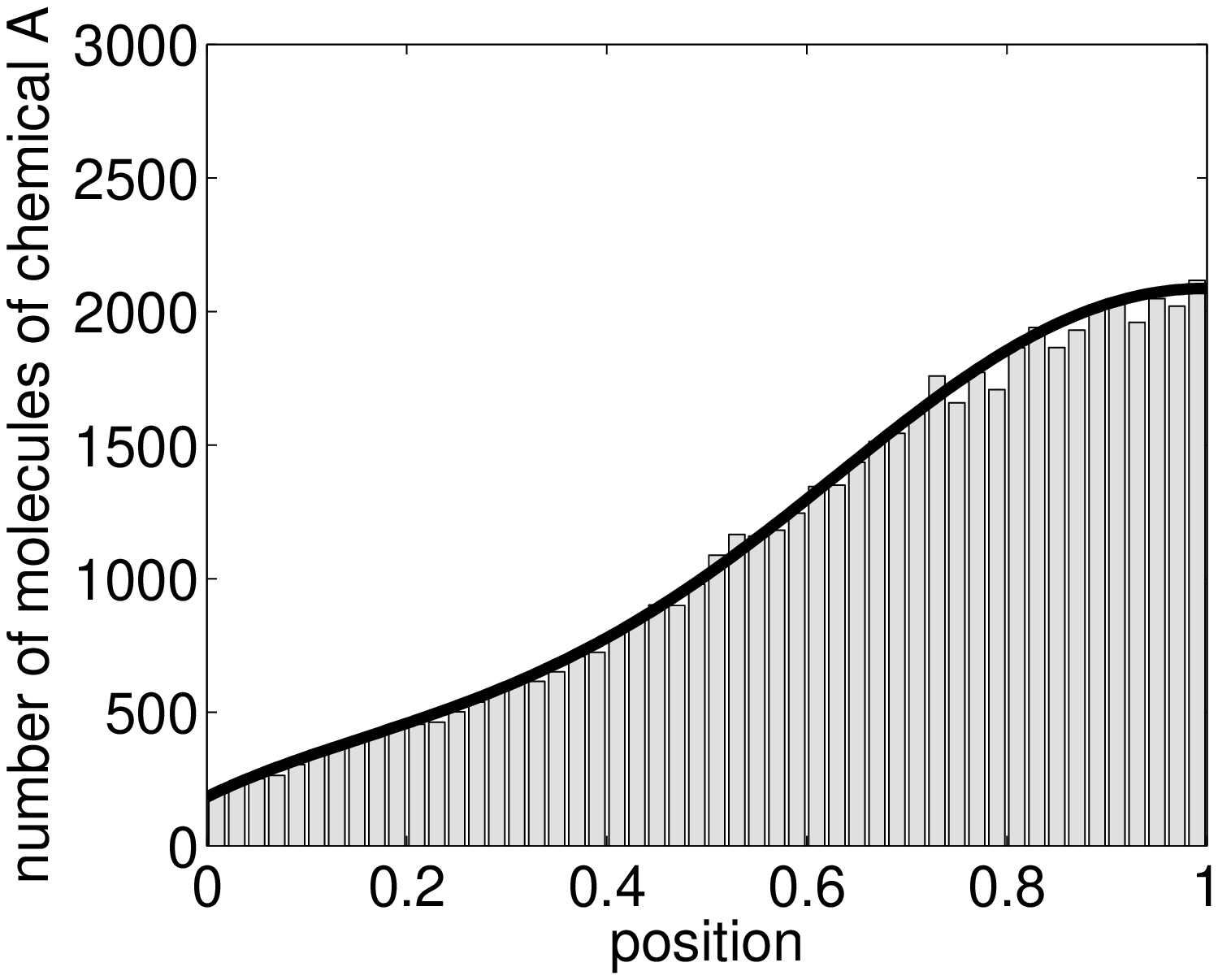,height=2.45in}
$\!\!\!\!\!$
\psfig{file=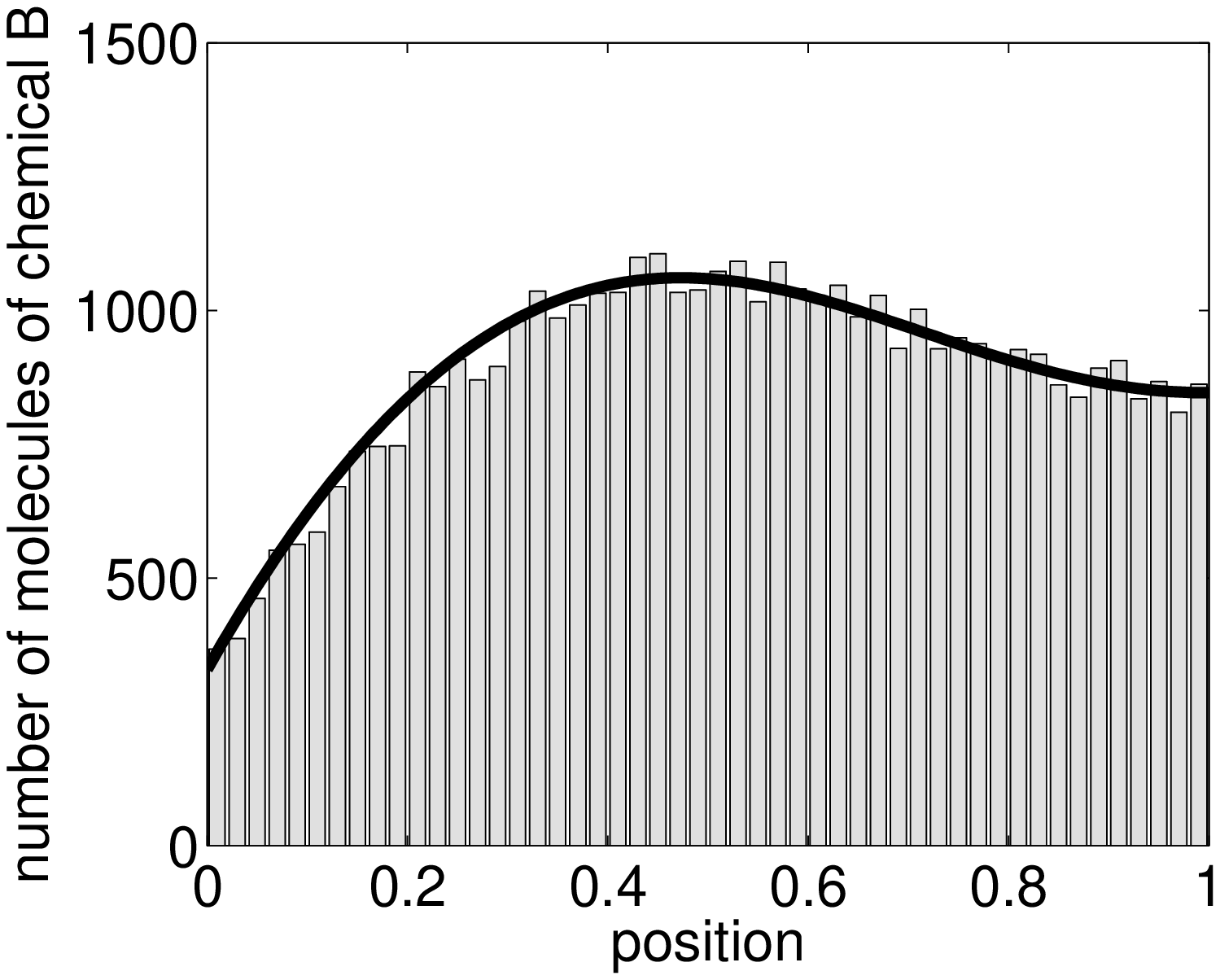,height=2.45in}
}
\caption{{\it Stochastic simulations of the reaction-diffusion
model of the Schnakenberg kinetics with the partially
adsorbing boundary at $x=0$ (histograms). Panel on
the left shows chemical $A$ and panel on the right
chemical $B$. Solution of $(\ref{rd1})$-$(\ref{rd2})$ 
with the Robin boundary condition
$(\ref{abRobinBD})$ at $x=0$ and no-flux boundary condition
at $x=1$ is plotted as the solid line.}} 
\label{figRD1}
\end{figure}

We choose $M=50$ in what follows, i.e. $h=0.02$. Initially,
we have $\omega = 1000$ molecules of $A$ and $B$ in each 
compartment, i.e. $A_i = B_i = \omega$, $i=1, 2, \dots, M$,
at time $t=0$. The rate of production
of $A$ is $2 \omega$ molecules per compartment
per unit of time. The rate of production of
$B$ is $8 \omega$ molecules per compartment
per unit of time. The degradation rate of $A$ in
the $i$-th compartment is proportional to $6 A_i$
and $k_c$ is chosen to be $3 \omega^{-2}$.
Diffusion constants are $D_A = 1$ and $D_B=0.1$. 
We implement the following boundary condition
at $x=0$: {\it whenever a molecule of chemical $A$
(resp. $B$) hits the boundary, it is 
adsorbed with probability $P^A_1 h$
(resp. $P^B_1 h$), and reflected
otherwise}. We choose $P^A_1 = P^B_1 = 10$.
We consider the reflective boundary 
condition for both chemicals at right boundary
$x=1$. Number of molecules in each compartment
at time $t=1$ are plotted in Figure \ref{figRD1}
(histograms).

Since $\omega$ is chosen sufficiently large, we can compare 
the results of stochastic simulation with $\overline{A}=\omega a$
and $\overline{B}=\omega b$ where $a$, $b$ are the solution
of the system of reaction-diffusion equations
\begin{equation}
\frac{\partial a}{\partial t}
=
D_A \, \frac{\partial^2 a}{\partial x^2}
+ 2 - 6 a + 3 a^2 \, b
\label{rd1}
\end{equation}
\begin{equation}
\frac{\partial b}{\partial t}
=
D_B \, \frac{\partial^2 b}{\partial x^2}
+ 8 -  3 a^2 \, b
\label{rd2}
\end{equation}
with Robin boundary conditions at $x=0$ given by (\ref{Kmod1}), namely
\begin{equation}
\frac{\partial a}{\partial x} (0,t)
=
P^A_1 \, a(0,t),
\qquad
\frac{\partial b}{\partial x} (0,t)
=
P^B_1 \, b(0,t),
\label{abRobinBD}
\end{equation}
and with no-flux boundary conditions at right boundary $x=1$. 
The curves $\overline{A}=\omega a$ and $\overline{B}=\omega b$
at time $t=1$ are plotted in Figure \ref{figRD1} as solid lines 
for comparison. We see that the Robin boundary (\ref{Kmod1})
which was derived for the corresponding diffusion model gives
good results for the full reaction-diffusion simulation
as well.

We  note that  the system 
(\ref{rd1})-(\ref{rd2}) possesses a so-called Turing instability 
\cite{Murray:2002:MB} if the values of the
diffusion constants $D_A$ and $D_B$ are chosen  to be sufficiently
different. Our choice $D_A=1$ and $D_B=0.1$ falls in the regime
in which the homogeneous solution
$a= 10/6$ and $b=48/50$ of (\ref{rd1})-(\ref{rd2})
is stable. On increasing the ratio $D_A/D_B$ Turing
patterns would develop, and we would observe solutions
with multiple peaks (provided that the domain size 
is sufficiently large); see e.g. \cite{Qiao:2006:SDS}.

\subsection{Spatially localized reactions}

\label{secposjump2RD}

In some morphogenesis applications \cite{Shimmi:2005:FTD,Reeves:2005:CAE},
one assumes that some prepatterning in the domain exists
and one wants to validate the reaction-diffusion
mechanism of the next stage of the patterning
of the embryo. In this section, we present an
example motivated by this approach. We consider
one-dimensional domain $[0,3]$ and we use the molecular based
approach of Andrews and Bray \cite{Andrews:2004:SSC}, 
i.e. the diffusion model is given by Position Jump Process II.
We choose a small simulation time step $\Delta t$.
At each time step, a molecule is released at random points
in the subinterval $[1,2]$ with probability
$k_p \Delta t \ll 1$. Moreover, we assume that any
molecule is degraded with probability $k_d \Delta t \ll 1$
during the simulation time step. Here, $k_p$ and
$k_d$ are given constants.
We implement the following boundary condition 
at $x=0$: {\it whenever a molecule 
hits the boundary, it is 
adsorbed with probability $P_2 \sqrt{\Delta t}$, 
and reflected otherwise}. We consider the reflective 
boundary condition at right boundary $x=3$.

\begin{figure}
\centerline{
\psfig{file=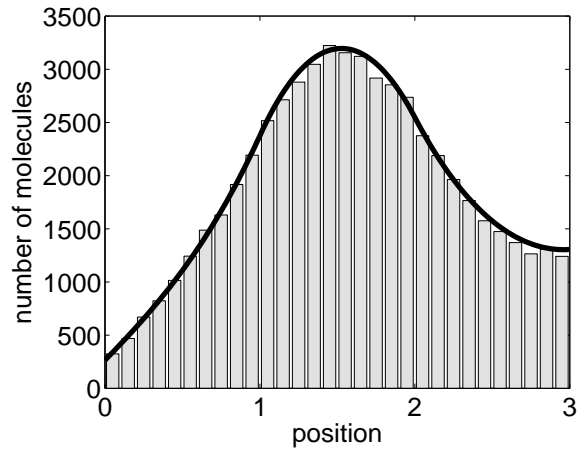,height=2.45in}
}
\caption{{\it Stochastic simulations of the reaction-diffusion
model with spatially localized reaction with the partially
adsorbing boundary at $x=0$ (histogram). 
Solution of $(\ref{rd3})$ with Robin boundary condition 
$(\ref{RobinBD})$, with $K\doteq 5.642$, and with no-flux boundary condition
at $x=3$, is plotted as the solid line.}} 
\label{figRD2}
\end{figure}

We start with no molecules in computational
domain $[0,3]$ at time $t=0$. We choose diffusion constant $D=1$,
time step $\Delta t=10^{-7}$, production
rate $k_p = 10^{5}$, degradation rate $k_d = 1$
and adsorption probability constant $P_2=5$.
To visualize the results, we divide the interval $[0,3]$
into 30 bins of length $0.1$ and we plot the number
of molecules in each bin at time $t=1$ in Figure \ref{figRD2}
(histogram). The results of the stochastic simulation
can be compared with the solution of reaction-diffusion
equation
\begin{equation}
\frac{\partial n}{\partial t}
=
D \, \frac{\partial^2 n}{\partial x^2}
+ 0.1 \, k_p \chi_{[1,2]}- k_d n,
\label{rd3}
\end{equation}
where $\chi_{[1,2]}$ is characteristic function of the interval
$[1,2]$. Equation (\ref{rd3}) is solved in the interval $[0,3]$ 
together with Robin boundary condition (\ref{RobinBD}), with $K$ 
given by (\ref{Kmod2}), and with no-flux boundary condition
at $x=3$. Using $P_2=5$, formula (\ref{Kmod2}) implies
that $K \doteq 5.642$. The density profile $n(x,1)$ at time $t=1$
is plotted in Figure \ref{figRD2} as a solid line 
for comparison. We see that the Robin boundary condition
(\ref{Kmod2}), which was derived for the corresponding diffusion 
model, gives good results for the full reaction-diffusion simulation
algorithm too.

\subsection{Mathematical justification}

\label{secmathRD}

Let us consider a system  of $k$ chemicals diffusing and reacting
in domain $[0,L]$. Let us suppose that the diffusion model is
the Position Jump Process I, i.e. we use the stochastic approach
based on the reaction-diffusion master 
equation \cite{Hattne:2005:SRD, Isaacson:2006:IDC} as in
Section \ref{secposjump1RD}. Let $p_1^j(t)$ (resp. $p_2^j(t)$)  
be the number of molecules of the $j$-th chemical, $j=1,2,\dots,k$,
at the boundary mesh point $x_1=h/2$ (resp. at $x_2=3h/2$).
If there are no reactions going on, then $p_1^j(t)$ and
$p_2^j(t)$ are related according to (\ref{p1equation}).
Introducing reactions at mesh point $x=h/2$, the boundary equation 
(\ref{p1equation}) is modified as follows
$$
p_1^j(t + \Delta t)
=
\left(1 - \frac{2 D \Delta t}{h^2} \right) p_1^j(t)
+
\frac{D \Delta t}{h^2}
\, \big(
p_{2}^j(t)
+
(1 - P_1 h) 
p_{1}^j(t)
\big)
+
\Delta t f(p_{1}^1(t),\dots,p_{1}^k(t))
$$
where $f(p_{1}^1(t),\dots,p_{1}^k(t))$ is the sum
of the rates of all reactions which modifies the $j$-th
chemical. Following the same procedure as in Section \ref{secmod1math},
we find out that the additional term does not influence
the boundary condition (it is $O(\Delta t)$ and only 
$O(\sqrt{\Delta t})$ terms have nonzero contribution to the
Robin boundary condition). In particular, we conclude
that the Robin boundary condition of the stochastic reaction-diffusion
model is given by (\ref{Kmod1}).

Next, let us consider the stochastic reaction-diffusion
model of Andrews and Bray \cite{Andrews:2004:SSC}
which was used in Section \ref{secposjump2RD}. Let
$p^j_{\Delta t} \equiv p^j_{\Delta t}(x,t): [0, \infty) 
\times \Delta t \, \en_0 \to [0, \infty)$ be the density function 
of the $j$-th chemical species, so that 
$p^j_{\Delta t}(x, i\Delta t) \dx$ is the number
of $j$-th molecules in interval $[x, x + \dx]$ 
at time $t=i\Delta t.$ If there are no reactions 
going on, then $p^j_{\Delta t}$ satisfies
the formula (\ref{onestepprob}). Introducing
the reactions to the system, formula (\ref{onestepprob})
is modified as follows:
\begin{equation}
p_{\Delta t}(x,t + \Delta t)
=
\int_0^\infty
[1 + O(\Delta t)]\,
p(x,t+\Delta t\,|\,y,t) \,
p_{\Delta t}(y,t)
\,\dy
\label{onestepprobRD}
\end{equation}
where the additional $O(\Delta t)$ term corresponds to the
reactions in the solution. As before, this term is of lower
order (compared to $O(\sqrt{\Delta t})$ terms) and
does not influence the Robin boundary condition. Consequently,
the Robin boundary condition (\ref{RobinBD}) is obtained
with $K$ given by (\ref{Kmod2}).

In this paper, we did not use the velocity jump 
processes to simulate stochas\-tically the reaction-diffusion 
process. Velocity jump processes are generally more 
computationally intensive. However, they might be of use 
if one considers that only sufficiently fast molecules 
can actually react. Alternatively, one can use
the approach based on binding/unbinding radii,
as in the Andrews and Bray method \cite{Andrews:2004:SSC},
to incorporate higher order reactions to the velocity jump models
of molecular diffusion. In any case, the reactions
are adding again $O(\Delta t)$ terms and do not influence
the boundary conditions, i.e. the boundary conditions
of the stochastic reaction-diffusion models can be chosen
as in the corresponding model of the diffusion only.

\section{Conclusions and outlook}

\label{summaryoutlook}

We have derived  the correct boundary
conditions for a number of stochastic models of reaction-diffusion
processes. For each model, we related the (microscopic)
probability of adsorption on the boundary with the
(macroscopic) Robin (reactive) boundary condition
(\ref{RobinBD}). First, we studied several stochastic models 
of  diffusion. We showed that each model
is suitable for the description of the molecular diffusion
far from the boundary. Moreover, we derived
formulas (\ref{Kmod1}), (\ref{Kmod2}), (\ref{Kmod3})
and (\ref{Kmod4}) relating reactivity $K$ of the boundary
with the parameters of the stochastic simulation
algorithms. Then, we showed that the boundary
conditions of stochastic models of reaction-diffusion
processes are the same as for the corresponding model
of diffusion only. We studied the stochastic approaches
based on the reaction-diffusion master equation
\cite{Hattne:2005:SRD, Isaacson:2006:IDC} and on the
Smoluchowski equation \cite{Andrews:2004:SSC}.
The main conclusion of this work is that a modeller can use
any of the stochastic model of the diffusion from
Section \ref{sec4models}, provided that
the adsorption probability on the reactive boundary
is chosen according to the corresponding formula,
i.e. (\ref{Kmod1}), (\ref{Kmod2}), (\ref{Kmod3})
or (\ref{Kmod4}), which is model-dependent.

We also presented the mathematical derivation of key
formulas (\ref{Kmod1}), (\ref{Kmod2}), (\ref{Kmod3})
and (\ref{Kmod4}). We devoted the most space to the
derivation of formula (\ref{Kmod2}) which is (to our
knowledge) a new
mathematical result. Derivation of formulas (\ref{Kmod1}) 
and (\ref{Kmod3}) is simple and we included the mathematical
arguments for completeness. The last formula (\ref{Kmod4}) 
has already appeared in literature \cite{Naqvi:1982:RFP}, 
though our derivation is more systematic.

It is interesting to note that in some applications, the
reactivity of the boundary depends also on the geometrical
constraints on the boundary. The binding sites on the surface
(e.g. reactive groups or receptors) become full as the adsorption 
progresses. Moreover, attaching large molecules to a binding site
can sterically shield the neighbourghing binding sites on 
the surface. For example, in  \cite{Erban:2006:DPI,Erban:2006:CPS}
we studied the chemisorption of polymers where the
attachment of a long polymer molecule to the surface prevents
attachment of other reactive polymers next to it.
This steric shielding was modelled using 
random sequential adsorption \cite{Evans:1993:RCS}.
In these models, an adsorption of one molecule to the
surface is attempted per unit of time. To relate
the time scale of random sequential adsorption
algorithms with physical time, one should 
couple the  theory of reactive boundaries presented here 
with algorithms which take the additional geometrical 
constraints on the boundary into account.
This is an area of ongoing research \cite{Erban:2006:TSR}.
 
\section*{Acknowledgments}

This work was supported by the Biotechnology and Biological 
Sciences Research Council.

\section*{Appendix. Robin boundary condition and chemistry}

In this appendix, we show the relation between the Robin 
boundary condition (\ref{RobinBD}) and the experimentally measurable
chemical properties of the boundary. Let us consider a
chemical diffusing in the domain $[0,L]$ which is adsorbed
by boundary $x=0$ with some rate $\overline{K}$. This
problem can be described by the reaction-diffusion equation
\begin{equation}
\frac{\partial n}{\partial t}
=
D \frac{\partial^2 n}{\partial x^2}
- \overline{K}  n \, \delta(x),
\qquad
\mbox{for} \; x \in [0,L], \; t \ge 0,
\label{diffusionequation1Dalt}
\end{equation}
together with no-flux boundary conditions, 
where $D$ is the diffusion constant and $\delta(x)$
a Dirac delta function. The term $\overline{K} n \, \delta(x)$ is a standard
description of reaction kinetics, localized on the
boundary. In the paper, we used an alternative
description of the chemically adsorbing boundary, 
given by the diffusion
equation (\ref{diffusionequation1D}) accompanied
by the Robin boundary condition (\ref{RobinBD}).
It is interesting to note that the constant $K$ in
(\ref{RobinBD}) is actually equal to the experimentally
measurable constant $\overline{K}$. To see this,
we discretize (\ref{diffusionequation1Dalt}) with space
step $h$ and we denote $n_0(t) = n(h/2,t)$ and $n_1(t) = n(3h/2,t)$.
Using a no-flux boundary condition (i.e. $n(-h/2,t)\equiv n(h/2,t)$), 
the discretization of (\ref{diffusionequation1Dalt}) gives
\begin{equation}
\frac{\partial n_0}{\partial t}
=
D
\frac{n_1 - n_0}{h^2}
-
\overline{K} n_0 \frac{1}{h} 
\label{discrete}
\end{equation}
which is equivalent to 
\begin{equation}
\frac{\partial n_0}{\partial t}
=
D
\frac{n_1 - n_0 + h \overline{K} n_0}{h^2}.
\label{discrete2}
\end{equation}
The same equation can be obtained by the discretization
of the diffusion equation (\ref{diffusionequation1D}) 
together with the Robin boundary condition (\ref{RobinBD}).
Hence we showed that $K = \overline{K}$, i.e. the
Robin boundary condition (\ref{RobinBD}) is indeed
the correct macroscopic description of the
chemically reacting boundary.

\section*{References}
\bibliographystyle{amsplain}
\bibliography{bibrad}
\end{document}